\documentstyle[onecolumn,graphics]{mn}
\newcommand{\etal}{{et al.}~}

\newcommand{\de}{\delta}
\newcommand{\p}{\partial}
\newcommand{\f}{\frac}

\newcommand{\Lam}{\Lambda}

\newcommand{\eps}{\epsilon}
\newcommand{\Om}{\Omega}

\newcommand{\s}{\sigma}
\newcommand{\al}{\alpha}
\newcommand{\fde}{\tilde{\delta}}
\newcommand{\fpsi}{\tilde{\psi}}

\newcommand{\fW}{\widetilde{W}}
\newcommand{\fT}{\widetilde{T}}
\newcommand{\bfx}{{\bf x}}
\newcommand{\bfr}{{\bf r}}

\newcommand{\bfk}{{\bf k}}

\newcommand{\bfq}{{\bf q}}
\newcommand{\bfg}{{\bf g}}
\newcommand{\bfp}{{\bf p}}
\newcommand{\bfu}{{\bf u}}
\newcommand{\bfL}{{\bf L}}
\newcommand{\bfS}{{\bf S}}
\newcommand{\bfT}{{\bf T}}

\newcommand{\calD}{{\cal D}}
\newcommand{\calE}{{\cal E}}
\newcommand{\calF}{{\cal F}}

\newcommand{\calJ}{{\cal J}}

\newcommand{\calT}{{\cal T}}
\newcommand{\bc}{\begin{center}}
\newcommand{\be}{\begin{equation}}
\newcommand{\ee}{\end{equation}}
\newcommand{\ec}{\end{center}}
\newcommand{\lan}{\langle}
\newcommand{\ran}{\rangle}
\title[Non-linear evolution of the tidal angular momentum]
{Non-linear evolution of the angular momentum
of protostructures from tidal torques} 

\author[P. Catelan and T. Theuns]
{Paolo Catelan and Tom Theuns\\
Department of Physics, Astrophysics, University of Oxford, Keble Road, 
Oxford OX1 3RH, UK \\} 

\begin{document}

\maketitle

\begin{abstract}
We discuss the non-linear evolution of the angular momentum $\bfL$
acquired by protostructures, like protogalaxies and protoclusters, due
to tidal interactions with the surrounding matter inhomogeneities. The
primordial density distribution is assumed to be Gaussian and the
non-linear dynamics of the collisionless mass fluid is followed using
Lagrangian perturbation theory. For a Cold Dark Matter spectrum, the
inclusion of the leading-order Lagrangian correction terms results in
a value of the rms ensemble average $\lan\bfL^2\ran^{1/2}$ which is
only a factor of 1.3 higher than the corresponding linear estimate,
irrespective of the scale. Consequently, the predictions of linear
theory are rather accurate in quantifying the evolution of the angular
momentum of protostructures before collapse sets in. In the Einstein-de
Sitter universe, the {\it initial} torque is a good estimate for
the tidal torque over the whole period during which the object is
spun up.
\end{abstract}


\section{Introduction}
The problem of the acquisition of angular momentum by protostructures
in the universe is of considerable interest in theories of galaxy and
cluster formation. A widely accepted view is that the present luminous
structures acquired their spin via gravitational tidal interactions
with the surrounding matter inhomogeneities (Hoyle 1949; Peebles 1969;
Doroshkevich 1970; White 1984; Barnes \& Efstathiou 1987; Hoffman
1986, 1988; Heavens \& Peacock 1988; Ryden 1988; Quinn \& Binney 1992;
Eisenstein \& Loeb 1995; Catelan \& Theuns 1996).  So far, however,
the theoretical analysis of the growth of the tidal galaxy angular
momentum has been essentially limited to the linear regime, during
which the galaxy spin grows proportionally to the cosmic time $t$
(Doroshkevich 1970; White 1984).

In this paper we examine analytically, for the first time, the
question of how the galaxy tidal angular momentum evolves during the
mildly non-linear regime. Previous attempts in this direction may be
found in Peebles (1969) and White (1984). However, in contrast to
their approach, we employ actual perturbative solutions of the
dynamical equations that describe the motion of the fluid. More
in detail, we apply Lagrangian perturbation theory.

The Lagrangian approach shows to be $ideal\,$ in treating the
evolution of the galaxy spin, because it is powerful in describing the
non-linear growth of the mass-density fluctuations on one hand
(Zel'dovich 1970a, b; Buchert 1992; Bouchet \etal 1992; Catelan 1995);
and on the other hand the usual difficulty of inverting the mapping
from Lagrangian coordinates $\bfq$ to Eulerian coordinates $\bfx$ is
completely by-passed. This is because the angular momentum $\bfL$ is
$invariant\,$ with respect to the Eulerian or Lagrangian description.

The layout of this paper is as follows: in the next section we first
briefly review the basics of the Lagrangian theory and the
perturbative solutions of the Lagrangian fluid equations. Next, we
compute within this framework the perturbative corrections to the
linear tidal angular momentum $\bfL^{(1)}$ acquired by a
protoobject. The resulting expressions are then simplified by
calculating their averages over the ensemble of realisations of the
linear Gaussian gravitational potential $\psi^{(1)}$ for objects with
given inertia tensor. We then compare the non-linear spin corrections
with the results of the linear analysis recently performed by Catelan
\& Theuns (1996). In the main text we restrict ourselves mainly to the
case of a flat universe, leaving the more involved treatment of closed
and open universes to Appendices.

\section{Non-linear evolution of the tidal angular momentum}
We consider a Newtonian pressureless and irrotational
self--gravitating fluid embedded in an expanding universe with
arbitrary density parameter $\Om$ but zero cosmological constant.
Such a fluid is assumed to mimic the behaviour of matter on scales
smaller than the horizon. Furthermore, we assume that luminous objects
like galaxies and clusters of galaxies started to grow due to
gravitational instability around primordial positive density
fluctuations $\de$ in this fluid.

We indicate by $\bfx$ comoving Eulerian coordinates, from which
physical distances may be obtained according to the law $\bfr =
a(t)\bfx$, where $a(t)$ is the expansion scale factor and $t$ the
standard cosmic time. We will use the temporal coordinate $\tau$ of
which the differential is defined by
\begin{equation}
d\tau \equiv a^{-2}\,dt\;,
\end{equation}
instead of $t$, since this allows a considerable simplification of the
formalism when dealing with the gravitational interactions in a
generic non-flat Friedmann universe (Shandarin 1980). The peculiar
velocity and the peculiar acceleration simplify when using $\tau$
instead of $t$ to:
\begin{equation} \f{d\bfx}{d\tau}\equiv
\dot{\bfx}\equiv a(\tau)\,\bfu(\bfx, \tau)\;,
\end{equation}
\begin{equation}
\f{d^2\bfx}{d\tau^2}\equiv \ddot{\bfx}\equiv \bfg(\bfx,\tau)\;.
\end{equation}
The dimensionless time $\tau$ is negative, as discussed in Shandarin
(1980), and the initial cosmological singularity at $t=0$ corresponds to
$\tau=-\infty$. In the open models, the infinity of the cosmic time,
$t=+\infty$, corresponds to $\tau=-1$; in the Einstein-de Sitter
universe, $t=+\infty$ corresponds to $\tau=0$; and in the closed
models the contraction phase starts at $\tau=0$. In terms of the
density parameter $\Om$, one has
\begin{equation}
\tau=-\sqrt{-k\,}\,(1-\Om)^{-1/2}\;,
\end{equation}
where $k$ is the curvature constant ($k=-1$ for open universes and
$k=1$ for closed universes). The case $\Om=1\,(k=0)$ is a singular
point for the transformation~(4) and in this case we take
$\tau\equiv-(3t)^{-1/3}$, which corresponds to using $a(t)\equiv
(3t)^{2/3}$ or $t_0=t/a^{3/2}\equiv 1/3$, which defines the unit of
time. The scale factor $a(\tau)$ may then be written for all Friedmann
models as $a(\tau)=(\tau^2+k)^{-1}$.

The linear evolution of angular momentum of protoobjects is most
easily analysed using the Zel'dovich (1970a, b) formulation (see White
1984; Catelan \& Theuns 1996) and the mildly non-linear spin growth is
most easily analysed using the Lagrangian perturbation theory. We
recall that the Zel'dovich approximation coincides with the linear
Lagrangian description.

We will essentially adopt the formulation of the Lagrangian
gravitational theory for a collisionless Newtonian fluid presented in
Catelan (1995; see also references therein) but note that in the
present paper, the variable $\tau$ has opposite sign and the growth
factor of the linear density perturbation is normalised
differently. An alternative formulation of the Lagrangian theory may
be found in Buchert (1992).

\subsection{Basic tools: Lagrangian theory}
In Lagrangian formulation, the departure of the mass elements
from the initial position $\bfq$ is described in terms of the 
displacement vector $\bfS$,
\begin{equation}
\bfx(\bfq, \tau) \equiv \bfq+\bfS(\bfq, \tau)\;.
\end{equation}
The trajectory $\bfx(\bfq, \tau)$ of the fluid element originally at
$\bfq$ satisfies the Lagrangian \lq irrotationality\rq~ condition and
the Poisson equation given by (Catelan 1995)
\begin{equation}
\eps_{\al\beta\gamma}\,x_{\beta\s}^C\,\dot{x}_{\gamma\s} = 0\;,
\label{eq:master1}
\end{equation}
\begin{equation}
x_{\al\beta}^C\,\ddot{x}_{\beta\al} = \al(\tau)[J-1]\;,
\label{eq:master2}
\end{equation}
respectively, where $\eps_{\al\beta\gamma}$ is the totally
antisymmetric Levi-Civita tensor of rank three, $\eps_{123}\equiv 1$,
and summation over repeated Greek indices (where $\al = 1, 2, 3$) is
understood. In these equations, $\al(\tau)\equiv 6a(\tau)$ and
$J\equiv1/(1+\de)$ is the determinant of the Jacobian of the mapping
$\bfx\rightarrow\bfq$. The determinant $J$ is non zero until the first
occurrence of shell-crossing (see, e.g., Shandarin \& Zel'dovich
1989). Furthermore, $x_{\al\beta}\equiv \p x_{\al}/\p q_{\beta}$ and
$x^C_{\al\beta}$ denotes the cofactor of $x_{\al\beta}$: we recall
that the latter is a quadratic function of $x_{\alpha\beta}$ and
consequently, the master equations~(\ref{eq:master1}) and
(\ref{eq:master2}) are cubic in $x_{\alpha\beta}$. In general,
$x_{\al\beta}$ is not a symmetric tensor: $x_{\al\beta}=x_{\beta\al}$
if, and only if, the Lagrangian motion is longitudinal in which case
$\bfx$ can be obtained from the gradient of a potential.

The irrotationality condition and the Lagrangian Poisson equation can
be written in terms of the displacement field $\bfS$ as:
\begin{equation}
\eps_{\al\beta\gamma}\,
\left[ (1+\nabla\cdot\bfS)\,\de_{\beta\sigma} - S_{\beta\sigma}
+ S^C_{\beta\sigma} \right] \dot{S}_{\gamma\sigma} = 0\;,
\end{equation}
\begin{equation}
\left[ (1+\nabla\cdot\bfS)\,\de_{\al\beta} - S_{\al\beta}
+ S^C_{\al\beta} \right] \ddot{S}_{\beta\al} 
=  \al(\tau)[J(\bfq,\tau)-1]\;,
\end{equation}
where $\nabla \equiv \nabla_{\bfq}$ and the symbol $\de_{\al\beta}$ indicates 
the Kronecker tensor. These equations are the closed set of general dynamical 
equations for the displacement vector $\bfS$ describing the motion of a 
collisionless fluid in the Lagrangian $\{\bfq\}$--space, embedded in an 
arbitrary Friedmann universe and subject to the Newtonian gravitational 
interaction of the mass fluctuations $J^{-1}-1$. We briefly summarise their 
perturbative solutions in the next subsection.

\subsection{Lagrangian perturbative solutions}
The exact Lagrangian equations~(8) and (9) are non-linear and non-local in 
the displacement $\bfS$ (see the discussion in Kofman \& Pogosyan 1995) and 
it is undoubtedly very difficult to solve them rigorously. A possible 
alternative is to seek for approximate solutions by expanding the 
trajectory $\bfS$ in a perturbative series, the leading term being the linear
displacement which corresponds to the Zel'dovich approximation: 
$\bfS=\bfS_1+\bfS_2+\bfS_3+\ldots$, where $\bfS_n=O(\bfS_1^n)$ is the $n$-th 
order approximation. Note that a perturbation series of this form
needs to include at least the third-order term $\bfS_3$ to capture the
essential physics contained in the cubic equations~(8)
and (9).

For the sake of simplicity we will limit ourselves to the Einstein-de
Sitter universe in the main text. In this universe,
$a=\tau^{-2}$. The reader interested in more general Friedmann models
is addressed to Appendix~A. We neglect decreasing modes.

\subsubsection{First-order approximation: Zel'dovich approximation}
The first-order solution to equations~(8) and (9) is separable in
space and time and corresponds to the Zel'dovich approximation
(Zel'dovich 1970a, b):
\begin{equation}
\bfS_1(\bfq, \tau)=D(\tau)\,\bfS^{(1)}(\bfq)\equiv 
D(\tau)\,\nabla\psi^{(1)}(\bfq)\;;
\end{equation}
the function $D(\tau)$ is the growth factor of linear density
perturbations and is given in the Einstein-de Sitter universe by:
\begin{equation}
D(\tau)=\tau^{-2}=a(\tau)\;.
\end{equation}
The expression of $D(\tau)$ valid in a generic Friedmann model is
reported in Appendix A. The function $\psi^{(1)}(\bfq)$ is the (initial)
gravitational potential. For later use, we define its Fourier transform,
$
\fpsi^{(1)}(\bfp)=
\int d\bfq\,\psi^{(1)}(\bfq)\,{\rm e}^{-i\bfp\cdot\bfq}\;,
$
where $\bfp$ is the comoving Lagrangian wave vector.
The Fourier transform of the linear density field, $\fde^{(1)}(\bfp, \tau)
=D(\tau)\fde_1(\bfp)$,
is related to $\fpsi^{(1)}(\bfp)$ via the Poisson equation, 
$
\fpsi^{(1)}(\bfp)=p^{-2}\fde_1(\bfp)\;.
$\\

\subsubsection{Second-order approximation}
The second-order solution is again separable with respect to the
spatial and temporal variables and describes a longitudinal motion in
Lagrangian space:
\begin{equation}
\bfS_2(\bfq, \tau)=E(\tau)\,\bfS^{(2)}(\bfq)\equiv 
E(\tau)\,\nabla\psi^{(2)}(\bfq)\;.
\end{equation}
The growing mode $E(\tau)$ is for the Einstein-de Sitter model given by:
\begin{equation}
E(\tau)=-\f{3}{7}\tau^{-4}\;.
\end{equation}
The analytic expression for $E(\tau)$ in a non-flat universe and
the expression for $\fpsi^{(2)}(\bfp)$ are reported in Appendix A.

\subsubsection{Third-order approximation}
The third-order solution $\bfS_3$ corresponds to three separable
modes, two longitudinal and one transverse, denoted by subscripts $a$,
$b$ and $c$, respectively:
\begin{equation}
\bfS_3(\bfq, \tau)=
F_a(\tau)\,\bfS_a^{(3)}(\bfq)+F_b(\tau)\,\bfS_b^{(3)}(\bfq)+
F_c(\tau)\,{\bf T}^{(3)}(\bfq) 
=
F_a(\tau)\,\nabla\psi_a^{(3)}(\bfq)+F_b(\tau)\,\nabla\psi_b^{(3)}(\bfq)+
F_c(\tau)\,\nabla\times{\bf A}^{(3)}(\bfq) \;.
\end{equation}
The growing modes $F_a, F_b$ and $F_c$ for a flat universe are, respectively,
\begin{equation}
F_a(\tau) = -\f{1}{3}\tau^{-6}\;,
\end{equation}
\begin{equation}
F_b(\tau) = +\f{10}{21}\tau^{-6}\;,
\end{equation}
\begin{equation}
F_c(\tau) = -\f{1}{7}\tau^{-6}\;.
\end{equation}
We stress that the transverse mode $\bfT^{(3)}$ does not describe any
physical vorticity in the fluid since the latter is assumed to be
irrotational. Rather, the occurrence of a transverse component is due
to the fact that a Lagrangian frame of reference is not
inertial. Consequently, this \lq fictitious\rq~ term is required to
obtain a correct physical description of the motion. Unfortunately,
this transverse mode is forgotten in some of the relevant literature
on the subject (e.g. Bouchet \etal 1995).

The general expressions for $F_a, F_b$ and $F_c$ for a non-flat
universe as well as the potentials $\fpsi_a^{(3)}(\bfp)$ and
$\fpsi_b^{(3)}(\bfp)$ and the transverse components $\fT_\al$ are
given in Appendix~A.\\

Comparing equations~(11), (13) and (15-17), it is clear that the
perturbative expansion for $\bfS$ is in fact a Taylor series in the
variable $D(\tau)=\tau^{-2}$. As the general expressions reported in
the Appendix A testify, this is no longer true in a non-flat
universe. However, it is shown there that the higher-order growth
factors can be approximated exceedingly well by powers of $D$:
$E\propto D^2$ and $F\propto D^3$, so the expansion is still \lq
close\rq~to a Taylor expansion. We will consider next how the
perturbative series in $\bfS$ translates into a perturbative series
for the angular momentum $\bfL$.

\subsection{Non-linear spin dynamics}
Let us consider some volume $V$ of the Eulerian $\bfx$-space. The
Cartesian coordinate system is assumed to be centred at the centre of
mass of $V$. Since we are interested in the intrinsic angular momentum
of the mass contained in $V$, we disregard the centre of mass motion.

The angular momentum $\bfL$ of the matter contained at the time $t$ in the 
volume $V$ is
\begin{equation}
\bfL(t) = \rho_b(t)\,a(t)^4
\int_{V(t)}d\bfx\,[1+\de(\bfx, t)]\,\bfx(t)\times\bfu(\bfx, t)=
\eta_0\,\int_\Gamma d\bfq\;[\bfq+\bfS(\bfq, \tau)]\times\f{d\bfS(\bfq, \tau)}
{d\tau}\;,
\end{equation}
where we have substituted the time variable $\tau$ in favour of the
standard cosmic time $t$ in the second integral. Here, the matter
density field is $\rho =\rho_b(1+\de)$, where $\rho_b(\tau)$ is the
background mean density and $\de$ is the density fluctuation field,
which is assumed to be initially Gaussian distributed and
$\eta_0\equiv a^3\rho_b$; the peculiar velocity field is denoted by
$\bfu$ (see, e.g., Peebles 1980).

The second equality in equation~(18) stresses the important fact that
the integral over the $Eulerian$ volume $V$ may be written equally
well as an integral over the corresponding (initial) $Lagrangian$
volume $\Gamma$. This enables us to apply the Lagrangian description
of the Newtonian gravity previously reviewed. The linear regime
(Zel'dovich approximation) has been fully analysed in this way by
Doroshkevich (1970), White (1984) and Catelan \& Theuns (1996),
whereas its Eulerian counterpart was studied by Heavens and Peacock
(1988). We can extend the Lagrangian analysis of the evolution of the
angular momentum $\bfL(\tau)$ to the non-linear regime by applying
perturbation theory to equation~(18). Perturbative corrections to
$\bfS(\bfq, \tau)$ (Bouchet \etal 1992; Buchert 1994; Catelan 1995 and
references therein) then give perturbative corrections to
$\bfL(\tau)$:
\begin{equation}
\bfL(\tau)=\eta_0\,\int_\Gamma d\bfq\;
\left[\sum_{i=0}^\infty\,\bfS_i(\bfq, \tau)\right]\times
\f{d}{d\tau}
\left[\sum_{j=0}^\infty\,\bfS_j(\bfq, \tau)\right]
\equiv
\sum_{h=0}^\infty\,\bfL^{(h)}(\tau)\;,
\end{equation}
where we have defined
\begin{equation}
\bfL^{(h)}(\tau)\equiv
\sum_{j=0}^h\,\eta_0\! \int_\Gamma d\bfq\, \,\bfS_j(\bfq, \tau)\times
\f{d \bfS_{h-j}(\bfq, \tau)}{d\tau}\;,
\end{equation}
with $\bfS_0\equiv\bfq$, hence $\bfL^{(0)}={\bf 0}$. Since we will be
interested in calculating second-order corrections to the ensemble
average $\lan \bfL^2\ran$, we need to compute corrections to $\bfL$ up
to third-order. After briefly reviewing the results of the linear
theory, we summarise the final expressions of the corrections
$\bfL^{(2)}$ and $\bfL^{(3)}$. The reader interested in the details of
the calculations is addressed to the Appendix B.

\subsubsection{Linear approximation}
The linear Lagrangian theory corresponds to the Zel'dovich
approximation and the first-order term in equation~(19) is given by:
\begin{equation}
\bfL^{(1)}(\tau)=
\eta_0\,\dot{D}(\tau)\int_{\Gamma}d\bfq\,\bfq\times\nabla\psi^{(1)}(\bfq)\;.
\end{equation}
If $\psi^{(1)}(\bfq)$ is adequately represented in the volume $\Gamma$
by the first three terms of the Taylor series about the origin
$\bfq={\bf 0}$, then each component $L^{(1)}_{\al}(t)$ may be written
in a compact form as (White 1984):
\begin{equation}
L^{(1)}_{\al}(\tau)=\dot{D}(\tau)\,\eps_{\al\beta\gamma}\,
\calD^{(1)}_{\beta\s}\,\calJ_{\s\gamma}\;,
\end{equation}
where we introduced the deformation tensor at the origin:
\begin{equation}
\calD^{(1)}_{\beta\s}\equiv\calD^{(1)}_{\beta\s}({\bf 0})=
\p_{\beta}\p_{\s}\psi^{(1)}({\bf 0})\;,
\end{equation}
and the inertia tensor of the mass contained in the volume $\Gamma$
\begin{equation}
\calJ_{\s\gamma}\equiv\eta_0\int_{\Gamma}d\bfq\,q_{\s}\,q_{\gamma}\;.
\end{equation}

Equation~(22) shows that the linear angular momentum $\bfL^{(1)}$ is
in general non--zero because the principal axes of the inertia tensor
$\calJ_{\al\beta}$, which depends only on the (irregular) shape of the
volume $\Gamma$, are not aligned with the principal axes of the
deformation tensor $\calD^{(1)}_{\al\beta}$, which depends on the
location of neighbour matter fluctuations. The temporal growth of
tidal angular momentum is completely contained in the function
$\dot{D}(\tau)$, which behaves as $\dot D(\tau)=-2\tau^{-3}\sim t$ in
the Einstein--de Sitter universe, as first noted by Doroshkevich
(1970). Finally, if $\Gamma$ is a spherical Lagrangian volume, then
$L^{(1)}_{\al}\sim\eps_{\al\beta\gamma}\,\calD^{(1)}_{\beta\gamma}=0$.
Consequently, the matter contained initially in a spherical volume
does not gain any tidal spin during the linear regime (see also the
discussion in White 1984).

\subsubsection{Second-order approximation}
The second-order term in equation~(19) involves the second-order
displacement $\bfS^{(2)}$:
\begin{equation}
\bfL^{(2)}(\tau)=\eta_0 \int_\Gamma d\bfq\,\bfq\times\f{d\bfS_2}{d\tau}=
\eta_0\,\dot{E}(\tau)\int_{\Gamma}d\bfq\,\bfq\times\nabla\psi^{(2)}(\bfq)\;.
\end{equation}
Note that the second-order term which follows from the product
$\bfS_1\times{d\bfS_1/d\tau}\,$ is identically zero since is involves
the product $\nabla\psi^{(1)}\times\nabla\psi^{(1)}$. The potential
$\psi^{(2)}$ is determined by the potential $\psi^{(1)}$ through the
equation~(64) in Appendix~A. Note that, since $E\propto\tau^{-4}$, one
has $\dot{E}\propto\tau^{-5}$ hence the second-order terms grows
$\propto t^{5/3}$ in the Einstein--de Sitter universe. This growth
rate was first derived by Peebles (1969). If we represent
$\psi^{(2)}(\bfq)$ in $\Gamma$ by the first three terms of a Taylor
series around $\bfq={\bf 0}$, as we did before for $\psi^{(1)}$, we
obtain for the $\al$-component:
\begin{equation}
L^{(2)}_{\al}(\tau)=\dot{E}(\tau)\,\eps_{\al\beta\gamma}\,
\calD^{(2)}_{\beta\s}\,\calJ_{\s\gamma}\;,
\end{equation}
where
\begin{equation}
\calD^{(2)}_{\beta\s}\equiv\calD^{(2)}_{\beta\s}({\bf 0})=
\p_{\beta}\p_{\s}\psi^{(2)}({\bf 0})\;.
\end{equation}
We can call $\calD^{(2)}_{\al\beta}$ the second-order deformation tensor.
The component $L^{(2)}_{\al}$ may be written in terms of the
second-order shear tensor $\calE^{(2)}_{\al\beta}$ as
\begin{equation}
L^{(2)}_{\al}(\tau)=\dot{E}(\tau)\,\eps_{\al\beta\gamma}\,
\calE^{(2)}_{\beta\s}\,\calJ_{\s\gamma}\;,
\end{equation}
where
\begin{equation}
\calE^{(2)}_{\beta\s}\equiv 
\calD^{(2)}_{\beta\s}-\f{1}{3}(\nabla\cdot\bfS^{(2)})\de_{\beta\s}
= (\p_{\beta}\p_{\s}-\f{1}{3}\de_{\beta\s}\nabla^2)\psi^{(2)}\;.
\end{equation}
An analogous relation is valid for the linear term (see Catelan \&
Theuns 1996). Note that the non-linear dynamical evolution modifies
only the deformation tensor and not the Lagrangian inertia
tensor. Furthermore, if $\Gamma$ is a sphere, then again
$\bfL^{(2)}={\bf 0}$. This is in contrast to Eulerian perturbation
theory, where the angular momentum of an Eulerian sphere does grow in
second-order perturbation theory (Peebles 1969; White 1984).

\subsubsection{Third-order approximation}
The two longitudinal modes $\bfS^{(3)}_a$, $\bfS^{(3)}_b$ and the
transverse one $\bfT^{(3)}$, and the coupling between first and second
order displacements originate the following third-order spin
corrections:
\begin{equation}
\bfL^{(3)}(\tau)=\eta_0 \int_\Gamma
d\bfq\,\bfq\times\f{d\bfS_3}{d\tau}
+\eta_0\int_\Gamma d\bfq\,
\left(\bfS_1\times\f{d\bfS_2}{d\tau}+\bfS_2\times\f{d\bfS_1}{d\tau}\right)=
\bfL_a^{(3)}(\tau)+\bfL_b^{(3)}(\tau)+\bfL_c^{(3)}(\tau)+\bfL^{(12)}(\tau)\;,
\end{equation}
where
\begin{equation}
\bfL_h^{(3)}(\tau)=
\eta_0\,\dot{F_h}(\tau)\int_{\Gamma}d\bfq\,\bfq\times\nabla\psi_h^{(3)}(\bfq)\;,
\end{equation}
\begin{equation}
\bfL_c^{(3)}(\tau)=
\eta_0\,\dot{F_c}(\tau)\int_{\Gamma}d\bfq\,\bfq\times\bfT^{(3)}(\bfq)\;,
\end{equation}
and $h=a$, $b$. Here, the temporal functions are such that
$\dot{F_a}\propto\dot{F_b} \propto\dot{F_c}\propto \tau^{-7}\propto
t^{7/3}$ in the Einstein-de Sitter universe. Proceeding as in the
previous cases by expanding the fields $\psi^{(3)}_a$, $\psi^{(3)}_b$
and $T^{(3)}_\al$ in Taylor series around $\bfq={\bf 0}$, one gets the
expressions,
\begin{equation}
L^{(3)}_{h\,\al}(\tau)=\dot{F_h}(\tau)\,\eps_{\al\beta\gamma}\,
\calD^{(3)}_{h\,\beta\s}\,\calJ_{\s\gamma}\;,
\end{equation}
\begin{equation}
L^{(3)}_{c\,\al}(\tau)=\dot{F_c}(\tau)\,\eps_{\al\beta\gamma}\,
T^{(3)}_{\gamma\s}\,\calJ_{\beta\s}\;.
\end{equation}
In these last equations, we defined the third-order deformation
tensors
\begin{equation}
\calD^{(3)}_{h\,\beta\s}\equiv\calD^{(3)}_{h\,\beta\s}({\bf 0})=
\p_{\beta}\p_{\s}\psi_h^{(3)}({\bf 0})\;,
\end{equation}
and again $h=a$, $b$ and $T^{(3)}_{\al\beta}\equiv \p T^{(3)}_\al/\p
q_\beta$.
In addition, one has:
\begin{equation}
\bfL^{(12)}(\tau) = \eta_0 [D(\tau)\dot E(\tau)-E(\tau)\dot D(\tau)]\,
\int_\Gamma d\bfq\,\nabla\psi^{(1)}\times\nabla\psi^{(2)}\;,
\end{equation}
where $(D\dot E-E\dot D)\propto \tau^{-7}\propto t^{7/3}$ in the
Einstein-de Sitter universe. Taylor expanding as before results in
\begin{equation}
L^{(12)}_\al=(D\dot{E}-E\dot{D})\,\eps_{\al\beta\gamma}
\Big[\eta_0\,\Gamma\,\p_\beta\psi^{(1)}({\bf 0})\,\p_\gamma\psi^{(2)}({\bf 0})
+\calD^{(1)}_{\beta\s}\,\calD^{(2)}_{\gamma\eta}\,\calJ_{\s\eta}\Big]\;.
\end{equation}
A more explicit expression for the contribution $L^{(12)}_\al$ in
terms of integrals in Fourier space may be found in the
Appendix~B. Note that if one considers the volume $\Gamma$ to be
centered on a peak of the underlying density distribution, as is
presumably appropriate when studying the formation of collapsing
objects, the first term is zero since $\psi^{(1)}$ is an extremum at
the origin in that case.\\

For a single collapsing region enclosed in a volume $\Gamma$ it is
enough to evaluate equation~(19) at the time of maximum expansion
$\tau_M$ to compute its angular momentum. After $\tau_M$ the angular
momentum essentially stops growing since the collapsed object is less
sensitive to tidal couplings (Peebles 1969). However it is more useful
to compute the mean angular momentum of the object averaged over an
ensemble of realisations of the gravitational potential random field
$\psi^{(1)}$: this is particularly important in order to compare the
theory against statistical results obtained from $N$--body simulations
or from observations. This programme is carried out in the next
section.

\section{Ensemble averages}
We simplify the previous results by considering the expectation value
over the ensemble of realisations of $\psi^{(1)}$ of the square of
$\bfL$, $\lan \bfL^2\ran_\psi\equiv\lan \bfL^2\ran$, for objects with
{\it preselected} inertia tensor. The underlying motivation for this
is that it gives the appropriate analytical estimate to compare
against numerical simulations that make use of the Hoffman-Ribak
algorithm to set up a constrained density field that contains an
object with given inertia tensor (Hoffman \& Ribak 1991; van de
Weygaert \& Bertschinger 1996). Moreover, the resulting expectation
value is still a good estimate for the angular momentum of Gaussian
peaks in case where the correlation between gravitational potential
field and inertia tensor -- neglected here -- can be properly taken
into account, at least in the linear regime (see the discussion in
Catelan \& Theuns 1996): the exact calculation in the non-linear
regime appears intractable analytically. In addition, the procedure
could give some insight into perturbative spin corrections in the case
of more generic primordial non-Gaussian statistics.

Taking into account the mildly non-linear corrections, one has as
leading terms:
\begin{equation}
\lan\bfL^2\ran = 
\lan\bfL^{(1)2}\ran +\lan\bfL^{(2)2}\ran +2\lan\bfL^{(1)}\!\cdot\bfL^{(3)}
\ran+o(\tau^{-11})\;.
\label{eq:lseries}
\end{equation}
Note that the term $\lan \bfL^{(1)}\!\cdot\bfL^{(2)}\ran$ is zero if
the linear potential is assumed to be Gaussian distributed, as it
involves an odd number of random fields. This term would represent the
$lowest\,$-order perturbative correction to the linear term for the
more general case of non-Gaussian statistics (work in progress).

\subsection{Perturbative corrections}

\subsubsection{Linear approximation}
The linear ensemble average $\lan\bfL^{(1)2}\ran$ is computed and
discussed extensively in Catelan \& Theuns (1996):
\begin{equation}
\lan\bfL^{(1)2}\ran=\f{2}{15}\dot{D}(\tau)^2(\mu_1^2-3\mu_2)\,\s(R)^2\;,
\label{eq:linam}
\end{equation}
where the quantity $\s(R)^2$ is the mass variance on the scale $R$,
explicitly given by $ \s(R)^2\equiv (2\pi^2)^{-1}\int_0^{\infty}
dp\,p^6\,P_{\psi}(p)\fW(pR)^2\;.$ In this expression, $\fW$ denotes a
filter applied to the input power spectrum to remove any ultra-violet
divergence. We will use the Gaussian smoothing function
$\fW(pR)=\exp(-p^2R^2/2)$.  Equation~(39) holds for any power spectrum
$\lan\fpsi^{(1)}(\bfp_1)\,\fpsi^{(1)}(\bfp_2)\ran_\psi\equiv
(2\pi)^3\,\de_D(\bfp_1+\bfp_2)\,P_{\psi}(p)$ and the value of
$\lan\bfL^{(1)2}\ran$ depends on the normalisation of the spectrum;
the symbol $\de_D$ indicates the Dirac function.  The general
expression~(\ref{eq:linam}) is $independent\,$ of the details of the
shape of the boundary surface of the volume $\Gamma$: it depends only
on the quantities $\mu_1$ and $\mu_2$, which are respectively the
first and the second invariant of the inertia tensor
$\calJ_{\al\beta}$. Specifically, denoting the eigenvalues of the
inertia tensor by $\iota_1$, $\iota_2$ and $\iota_3$, one has
\begin{equation}
\mu_1\equiv\iota_1+\iota_2+\iota_3\;
\end{equation}
and
\begin{equation}
\mu_2\equiv\iota_1\iota_2+\iota_1\iota_3+\iota_2\iota_3\;.
\end{equation}
For a spherical volume, $\iota_1=\iota_2=\iota_3$, hence
$\mu_1^2-3\mu_2=0$, as we stressed before. For any volume $\Gamma$
one has $\mu_1^2-3\mu_2 \ge 0$.

\subsubsection{Higher-order approximation: $\lan\bfL^{(2)2}\ran$}
The calculation of the term $\lan\bfL^{(2)2}\ran$ takes advantage of
the results of the second-order approximation. The final expression,
assuming a power-law spectrum $P_{\de}(p)=Ap^n=P_{\psi}\,p^4$, may be
written as:
\begin{equation}
\lan\bfL^{(2)2}\ran=\f{2}{15}\,
\dot{E}(\tau)^2(\mu_1^2-3\mu_2)\,\Sigma^{(2)}(R; n)\;,
\end{equation}
where the function $\Sigma^{(2)}$ depends on the smoothing scale $R$
and the normalisation of the spectrum $A$, as
\begin{equation}
\Sigma^{(2)}(R; n) \equiv\f{A^2}{(2\pi)^4} \int_0^\infty dp \,p^{n+2}\,
[\fW(pR)]^2\,
\Delta^{(2)}(p; n)\;,
\end{equation}
\begin{equation}
\Delta^{(2)}(p_1; n)\equiv 
\f{1}{2^{2n-1}}\int_0^\infty dp_2\,p_2^{n+2}\int_{-1}^{+1}
d\theta\, \left(1-\theta^2\right)^2 \,
\big[(p_1\,p_2^{-1}+p_2\,p_1^{-1})^2-4\theta^2\big]^{(n-4)/2}\;.
\end{equation}
Typically, these integrals have to be evaluated numerically. The
results for more physical power spectrum, like the Cold Dark Matter
(CDM) spectrum, are discussed in Appendix B, where the details of the
derivation of the equation~(42) are given as well. Surprisingly, the
average $\lan\bfL^{(2)2}\ran$ factorises the same invariant
$\mu_1^2-3\mu_2$ of the inertia tensor $\calJ$ as appeared in the
linear term (equation 39).

\subsubsection{Higher-order approximation: $\lan\bfL^{(1)}\!\cdot\bfL_h^{(3)}\ran$}
The calculation of the correction
$\lan\bfL^{(1)}\!\cdot\bfL_h^{(3)}\ran$ takes advantage of the results
of the third-order Lagrangian theory. The displacement $\bfS_a^{(3)}$
does not induce on average any higher-order correction to the angular
momentum, since it corresponds to radial motions of the fluid patches
(see Appendix~B for an explicit derivation),
\begin{equation}
\lan\bfL^{(1)}\!\cdot\bfL_a^{(3)}\ran = 0\;.
\end{equation}
Assuming again a scale-free power spectrum $P_{\psi}(p)=Ap^{n-4}$, the 
corrections due to the third-order displacements $\bfS_b^{(3)}$
and $\bfT^{(3)}$ are (see Appendix B for details):
\begin{equation}
\lan\bfL^{(1)}\!\cdot\bfL_b^{(3)}\ran =\f{2}{15}\,
\dot{D}(\tau)\,\dot{F_b}(\tau)(\mu_1^2-3\mu_2)\,\Sigma_b^{(3)}(R; n)\;,
\end{equation}
\begin{equation}
\lan\bfL^{(1)}\!\cdot\bfL_c^{(3)}\ran =-\f{2}{15}\,
\dot{D}(\tau)\,\dot{F_c}(\tau)(\mu_1^2-3\mu_2)\,\Sigma_c^{(3)}(R; n)\;,
\end{equation}
where the functions $\Sigma_b^{(3)}$ and $\Sigma_b^{(3)}$ are
respectively
\begin{equation}
\Sigma_b^{(3)}(R; n) \equiv\f{A^2}{(2\pi)^4} \int_0^\infty dp \,p^{n+3}\,[\fW(pR)]^2\,
\Delta_b^{(3)}(p; n)\;,
\end{equation}
\begin{equation}
\Sigma_c^{(3)}(R; n) \equiv\f{A^2}{(2\pi)^4} \int_0^\infty dp \,p^{n+4}\,[\fW(pR)]^2\,
\Delta_c^{(3)}(p; n)\;,
\end{equation}
and the integrands $\Delta_b^{(3)}$ and $\Delta_c^{(3)}$
\begin{equation}
\Delta_b^{(3)}(p_1; n)\equiv \int_0^\infty dp_2\,p_2^{n+1}\int_{-1}^{+1}
d\theta\, \left(1-\theta^2\right)^2 \,
\big[(p_1\,p_2^{-1}+p_2\,p_1^{-1})-2\theta\big]^{-1}\;,
\end{equation}
\begin{equation}
\Delta_c^{(3)}(p_1; n)\equiv \int_0^\infty dp_2\,p_2^n\int_{-1}^{+1}
d\theta\, \left(1-\theta^2\right)\,\big(\theta-p_2\,p_1^{-1}\big) \,
\big[(p_1\,p_2^{-1}+p_2\,p_1^{-1})-2\theta\big]^{-1}\;.
\end{equation}
Again, these integrals have to be calculated numerically. The results
for a more physical power spectrum are discussed in Appendix B. Note
that once more these averages factorise out the invariant
$\mu_1^2-3\mu_2$ of the inertia tensor.

\subsubsection{Higher-order approximation: $\lan\bfL^{(1)}\!\cdot\bfL^{(12)}\ran$}
The perturbative correction $\lan\bfL^{(1)}\!\cdot\bfL^{(12)}\ran$
originates from the coupling between first- and second-order
displacements. The calculation of this term is cumbersome and is
detailed in Appendix~B: we restrict ourselves to giving the final
result:
\begin{equation}
\lan\bfL^{(1)}\!\cdot\bfL^{(12)}\ran =\f{2}{15}\,
\dot{D}(\tau)\left[D(\tau)\dot E(\tau)-\dot D(\tau)E(\tau)\right]
(\mu_1^2-3\mu_2)\,\Sigma^{(12)}(R; n)\;,
\end{equation}
where
\begin{equation}
\Sigma^{(12)}(R; n) \equiv \f{15}{2}\!
\int\! {d\bfp_1\,d\bfp_2\over
(2\pi)^6}\,\fW(p_1R)\,\fW(p_2R)\,\fW(|\bfp_1-\bfp_2|R)\,
P_\psi(p_1)\,P_\psi(p_2)\,{\kappa^{(2)}(\bfp_1,\bfp_2)\over
(\bfp_1-\bfp_2)^2}\,
p_{1z}(\bfp_2-\bfp_1)_z\,[(\bfp_1\cdot\bfp_2)p^2_{2z}-\bfp_2^2p_{1z}p_{2z}]
\;,
\end{equation}
where $\kappa^{(2)}$ is the second-order kernel defined in Appendix~A.
This integral needs to be computed numerically both for scale-free and
CDM spectra. Note that as previously this pertubative correction term
to the spin is proportional to $\mu_1^2-3\mu_2$. This remarkable
property enables us to calculate the relative contribution of linear
and higher-order spin terms.\\

The time dependent growth factors of the various terms in
equation~(\ref{eq:lseries}) are illustrated in Fig.~1 and the results
of numerically integrating the momentum contributions $\Sigma$ for
both power law and the CDM spectrum are shown in Fig.~2 as a function
of the spectral index $n$. These wavevector integrations generally
diverge at small wavelengths for scale-free spectra: to obtain finite
expressions we have filtered out this ultra-violet divergence by
smoothing the integrals over $p_2$ (in equations~44, 50 and 51)
artificially with a Gaussian filter of width $0.5\times h^{-1}$ Mpc.
The CDM integrals, in contrast, are finite. The numerical values of
these momentum integrals will be used in the next section to estimate
the relative contributions of the higher-order terms to the linear
estimate of $\lan\bfL^2\ran$. The dependence of these relative
momentum contributions on the smoothing radius $R$ for a CDM spectrum
is shown in Fig.~3, from which it is clear that, although the various
$\Sigma$'s depend strongly on $R$, the normalised contributions
$\Sigma/\sigma^4$ are more weakly dependent on the smoothing scale.

\begin{figure}
\setlength{\unitlength}{1cm}
\centering
\begin{picture}(10,12)
\put(0.0,0.0){\includegraphics{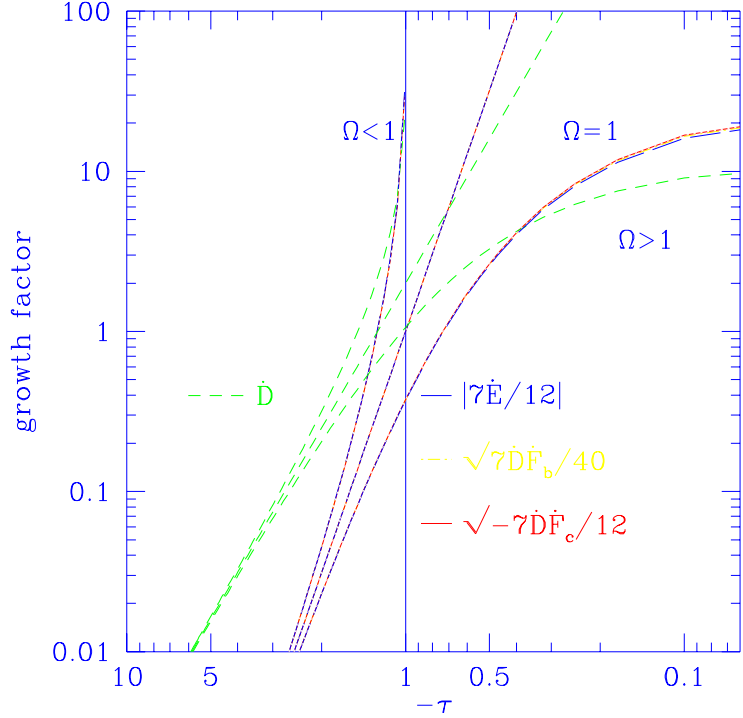}}
\end{picture}
\caption{Time dependencies, for open, flat and closed universes, (top,
middle and bottom curves respectively)
$\langle\bfL^{(1)2}\rangle^{1/2}\propto \dot D$,
$\langle\bfL^{(2)2}\rangle^{1/2}\propto |\dot E|$, $\langle
\bfL^{(1)}\!\cdot \bfL^{(3)}_b \rangle^{1/2}\propto \left(\dot D\dot
F_b \right)^{1/2}$,
$\langle\bfL^{(1)}\!\cdot\bfL^{(3)}_c\rangle^{1/2}\propto \left(-\dot
D\dot F_c\right)^{1/2}$ and
$\langle\bfL^{(1)}\!\cdot\bfL^{(12)}\rangle^{1/2}\propto \left(\dot D
(\dot D E-D\dot E)\right)^{1/2}$. The latter four have been scaled by
the indicated numerical factors to give the asymptotic behaviour
$(-\tau)^{-5}$ for $\tau\rightarrow-\infty$ and are practically
indistinguishable on the plot. The vertical line at $\tau=-1$ denotes
the infinity of physical time $t$ for open universes. The term
corresponding to $\dot D\dot F_a$ is not plotted since
$\langle\bfL^{(1)}\cdot\bfL^{(3)}_a\rangle=0\,$.}
\label{fig:factors}
\end{figure}

\begin{figure}
\setlength{\unitlength}{1cm}
\centering
\begin{picture}(10,12)
\put(0.0,0.0){\includegraphics{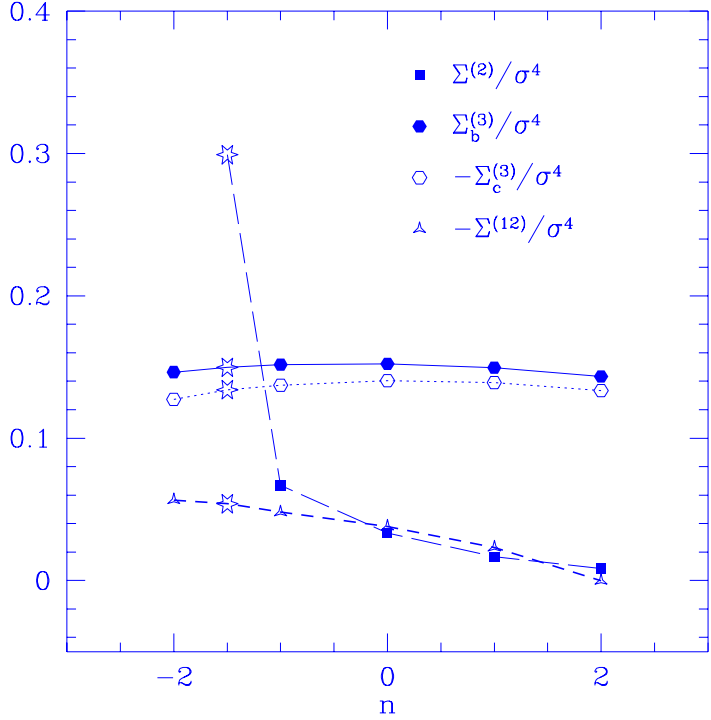}}
\end{picture}
\caption{Wave vector parts of higher-order corrections to
$\lan\bfL^2\ran$ computed for power law spectra $P_\psi(p)\propto
p^n$, filtered with a Gaussian smoothing function at $R=0.5\,h^{-1}\,$
Mpc, as a function of spectral index $n$, in units of the square of
the mass variance $\sigma^2$ in order to eliminate the dependence on
the normalisation of the spectrum. The points corresponding to the CDM
spectrum are indicated by stars and positioned arbitrarily at $n=-1.5$
which resembles the slope of CDM power spectrum at galactic scales.}
\label{fig:spatial}
\end{figure}

\section{Angular momentum at the maximum expansion time}
In this section we quantify the relative perturbative corrections
to the linear angular momentum by computing:
\begin{equation}
\Upsilon\equiv {\lan \bfL^2\ran\over\lan \bfL^{(1)^2}\ran} -1 = {\lan
\bfL^{(2)2}\ran\over\lan \bfL^{(1)^2}\ran} + 2\,{\lan \bfL^{(1)}
\!\cdot \bfL^{(3)} \ran\over\lan \bfL^{(1)^2}\ran} 
= \Upsilon^{(22)} + \Upsilon^{(13)}_a + \Upsilon^{(13)}_b 
+ \Upsilon^{(13)}_c + \Upsilon^{(112)}\;,
\end{equation}
where we recall that
$\bfL^{(3)}=\bfL^{(3)}_a+\bfL^{(3)}_b+\bfL^{(3)}_c+\bfL^{(12)}$ hence
$\bfL^{(3)}$ gives rise to four correction terms. From equation~(45)
we find immediately that $\Upsilon^{(13)}_a=0$. We compute the other
correction terms at the time defined by $D(\tau_M)\sigma(M)=1$ on the
mass scale $M$, which is close to the maximum expansion time found
from extrapolating the spherical model (e.g., Peebles 1980). After
$\tau_M$, the protoobject starts collapsing and tidal torques are much
less efficient in spinning up its matter content (Peebles 1969; Barnes
\& Efstathiou 1987). Therefore, we assume that the growth of the spin
effectively ceases after maximum expansion of the object and identify
the angular momentum at that time with the \lq final\rq~ angular
momentum. We understand that this is a partial description of the real
world (see the discussion in Catelan \& Theuns 1996).

Collecting the expressions for the various corrections, we find:
\begin{eqnarray}
\Upsilon^{(22)} &=& \left({\dot E\over D\dot D}\right)^2\,
{\Sigma^{(2)}(M)\over \sigma(M)^4}\,[D(\tau_M)\,\sigma(M)]^2 = 0.22\\
\Upsilon^{(13)}_b &=& 2\,{\dot F_b\over D^2\dot D}\,
{\Sigma^{(3)}_b(M)\over \sigma(M)^4}\,[D(\tau_M)\,\sigma(M)]^2 = 0.44\\
\Upsilon^{(13)}_c &=& -2\,{\dot F_c\over D^2\dot D}\,
{\Sigma^{(3)}_c(M)\over \sigma(M)^4}\,[D(\tau_M)\,\sigma(M)]^2 =-0.12\\
\Upsilon^{(112)} &=& 2\,{D\dot E-\dot D E \over D^2\dot D}\,
{\Sigma^{(12)}(M)\over \sigma(M)^4}\,[D(\tau_M)\,\sigma(M)]^2 =0.04\;,
\end{eqnarray}
where the numerical values are calculated for the (flat) standard CDM
model when filtered on the scale of $R=0.5\,h^{-1}$ Mpc and at the
maximum expansion time, i.e., when $D\,\sigma=1$. The factors $\dot
E/D\dot D$, $\dot F_b/D^2\dot D$, $\dot F_c/D^2\dot D$ and $(D\dot
E-\dot D E)/D^2\dot D$ do not depend on $\tau$ for a flat universe;
for a non-flat universe, their $\tau$ dependence is extremely weak, in
view of the excellent approximations $E\propto D^2$, $F_{b,c}\propto
D^3$ and $D\dot E-\dot D E\propto D^3$, as illustrated in Fig.~5. Note
that, with these approximations, the various $\Upsilon$'s do not
depend on the normalisation of the power spectrum.

\begin{figure}
\setlength{\unitlength}{1cm}
\centering
\begin{picture}(10,12)
\put(0.0,0.0){\includegraphics{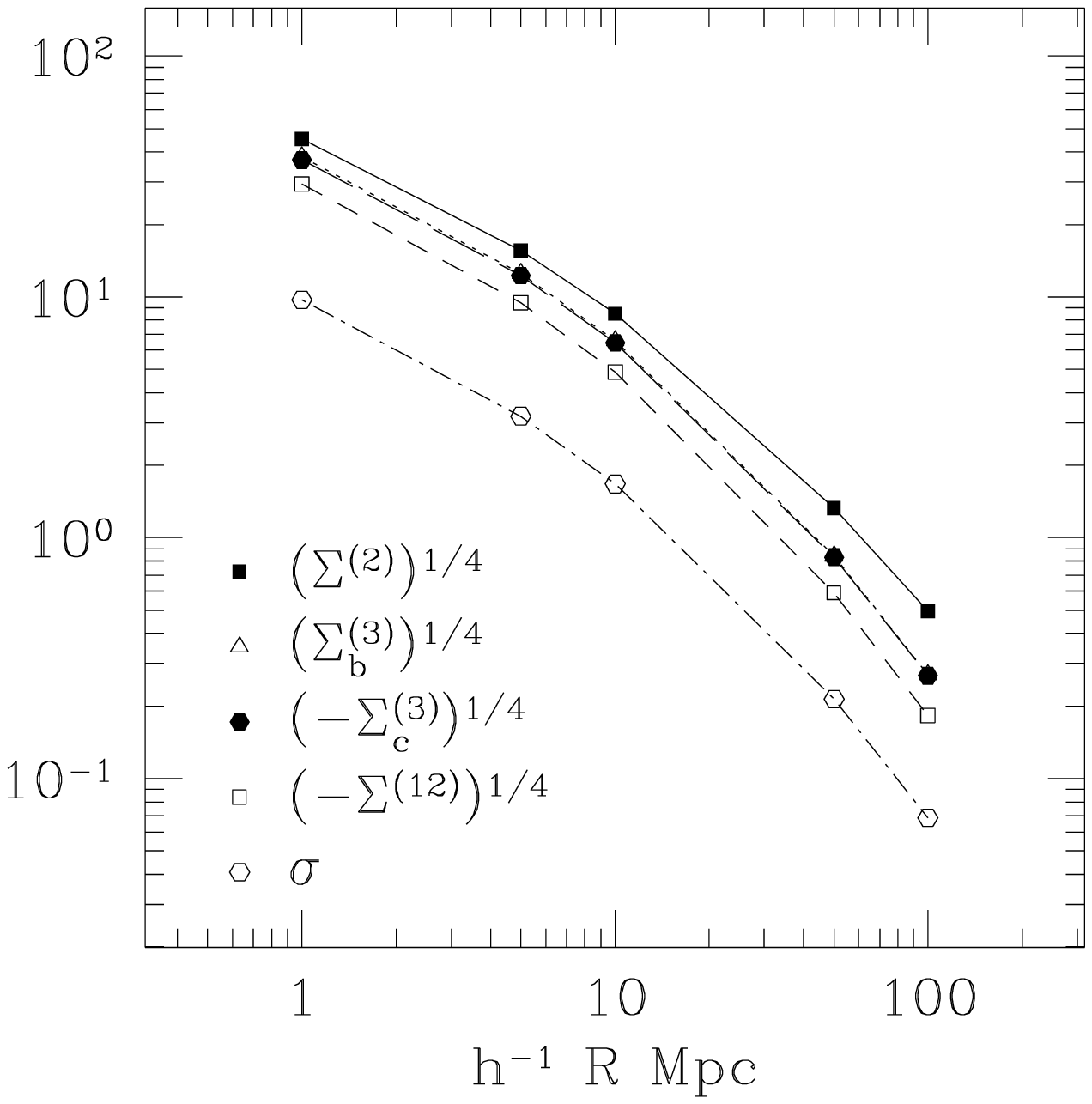}}
\put(0.0,0.0){\includegraphics{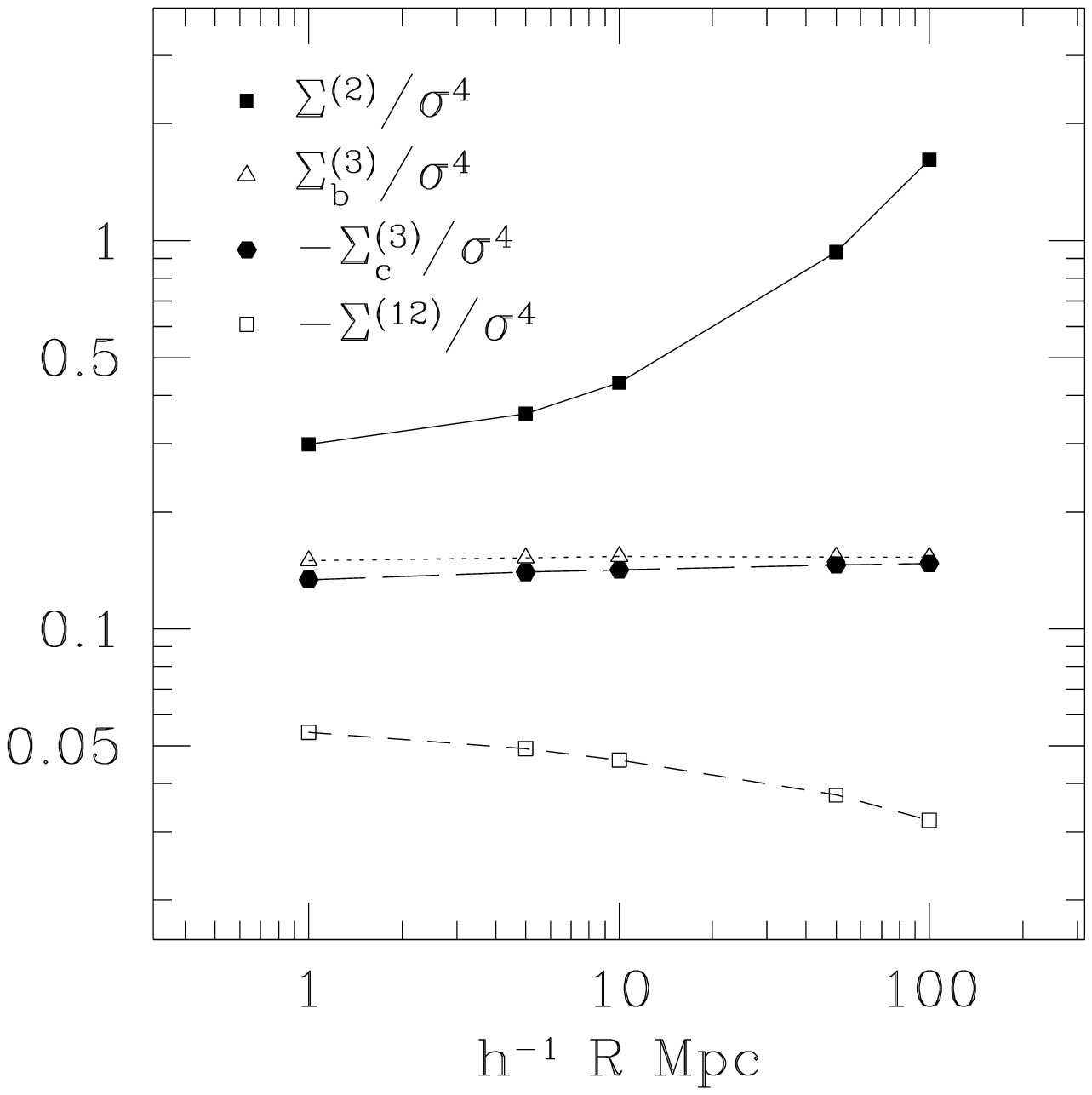}}
\end{picture}
\caption{Wave vector parts of non-linear corrections to
$\lan\bfL^2\ran$ for a standard CDM spectrum as a function of
filtering scale $R$. Left panel: scaling of $\Sigma^{(2)}$,
$\Sigma_b^{(3)}$, $\Sigma_c^{(3)}$ and $\Sigma^{(12)}$ versus $R$,
arbitrarily normalised. The run of the mass rms $\sigma$ with $R$ is
shown for comparison (same arbitrary normalisation). The values for
$\Sigma^{(3)}_b$ and $\Sigma^{(3)}_c$ are practically
indistinguishable on this plot. Right panel: the various $\Sigma$'s
normalised to $\sigma^4$ as a function of filtering scale $R$. These
ratios are independent of the normalisation of the power spectrum.}
\label{fig:cdmspec}
\end{figure}

We conclude that, since at maximum expansion
\begin{equation}
\lan\bfL^2\ran = (1+\Upsilon)\,\lan\bfL^{(1)2}\ran \approx 1.6\,\,
\lan\bfL^{(1)2}\ran\;,
\end{equation}
the linear estimate of $\lan \bfL^2\ran$ is roughly a factor 1.6 times
{\it smaller} than the value obtained when taking into account the
lowest-order non-linear corrections, for a standard CDM spectrum.
Hence, $\sqrt{\lan \bfL^2\ran}\approx 1.3\, \sqrt{\lan
\bfL^{(1)2}\ran}$. From this we conclude that the predictions of
linear theory are surprisingly accurate and that the dynamical
perturbative corrections appear converged.

\section{Summary and conclusions}
In this paper we analysed the growth of the tidal angular momentum
$\bfL$ acquired by a protoobject (protogalaxy or protocluster) during
the mildly non-linear evolution of the matter density perturbations,
assuming the latter to be Gaussian distributed. The dynamics of the
collisionless matter fluid is described using the Lagrangian approach
in the formulation given by Catelan (1995). This formulation is very
suitable to study the problem at hand, because the Lagrangian
expressions are considerably simpler than their Eulerian counterparts,
yet the protogalaxy's tidal spin is a vector {\it invariant} under the
change of Eulerian to Lagrangian spatial coordinates, $\bfx$ and
$\bfq$ respectively. Specifically, the difficult problem of inverting
the mapping $\bfx=\bfq+\bfS$, where $\bfS$ is the displacement vector,
in order to recover the Eulerian quantities from the Lagrangian ones,
is completely avoided.

The strategy we follow is straightforward. The non-linear spin
corrections $\bfL^{(h)}$, where $\bfL^{(1)}$ is the linear angular
momentum, are calculated approximating the fluid elements'
trajectories $\bfS$ by the perturbative solutions $\bfS_h$ of the
Lagrangian fluid equations~(8) and (9). This leads to the
expression~(20). Since we are interested in computing the lowest-order
perturbative corrections to the ensemble average $\lan\bfL^{(1)2}\ran$
for objects with given inertia tensor, we need to calculate
corrections to $\bfL$ up to third-order. This has the added advantage
that we take account of the full physical content of equations~(8) and
(9), since the latter are cubic in the displacement. The calculation
is summarised as follows: from the knowledge of
$\bfS=\bfS_1+\bfS_2+\bfS_3$ (where $\bfS_1$ corresponds to the
displacement in Zel'dovich approximation), we deduce the corresponding
corrections $\bfL=\bfL^{(1)}+\bfL^{(2)}+\bfL^{(3)}$ and finally get
the perturbative expansion $\lan\bfL^2\ran=
\lan\bfL^{(1)2}\ran+\lan\bfL^{(2)2}\ran+2\,\lan\bfL^{(1)}\!\cdot\bfL^{(3)}\ran$.
The term $\lan\bfL^{(1)}\!\cdot\bfL^{(2)}\ran$ is zero for an
underlying Gaussian matter distribution, but it should be taken into
account in the framework of more general non-Gaussian statistics (work
in progress). Assuming Gaussian statistics here, we disregard it. In
sections 2.1 and 2.2 (for the Einstein-de Sitter universe; in Appendix
A for a more general Friedmann universe) we reviewed the Lagrangian
theory and the perturbative solutions $\bfS_1, \bfS_2$ and $\bfS_3$ of
the Lagrangian fluid equations. Using these results we calculate the
corrections $\lan\bfL^{(2)2}\ran$ and
$\lan\bfL^{(1)}\!\cdot\bfL^{(3)}\ran$ in section 3, after summarising
the results of linear theory (i.e., the term
$\lan\bfL^{(1)2}\ran$). The final expressions are rather cumbersome
(the details of the calculations have been deferred to Appendix B),
but we can summarise the main features of our results as follows: for
an Einstein-de Sitter universe,
\begin{itemize}
\item $\lan\bfL^{(1)2}\ran^{1/2} \propto \tau^{-3} \propto t
\;\;\;\;\;\;\;\;\;\;\;\;\;
\;\;\;\;\;\;\;\;\;\;\;\;\;\left[\propto \dot{D}(\tau)\right]\;;$\\
\item $\lan\bfL^{(2)2}\ran^{1/2} \propto \tau^{-5} \propto t^{5/3}
\;\;\;\;\;\;\;\;\;\;\;\;\;\;
\;\;\;\;\;\;\;\left[\propto \dot{E}(\tau)\right]\;;$\\
\item $\lan\bfL^{(1)}\!\cdot\bfL_h^{(3)}\ran^{1/2} \propto \tau^{-5}
\propto t^{5/3} \;\;\;\;\;\;\;\;\;\;\;\;\;\;\left[\propto
\left(\dot{D}(\tau)\dot{F_h}(\tau)\right)^{1/2}\right]\;;$\\
\item $\lan\bfL^{(1)}\!\cdot\bfL^{(12)}\ran^{1/2} \propto \tau^{-5} \propto t^{5/3}
\;\;\;\;\;\;\;\;\;\;\;\;\;\left[\propto
\left(\dot{D}(\tau)[\dot D(\tau)E(\tau)-D(\tau)\dot E(\tau)]\right)^{1/2}\right]\;,$
\end{itemize}
\noindent where $D$ is the growth factor of the density perturbations,
$E$ and $F_h$ ($h=a,b,c$) are the growing modes of the second- and
third-order Lagrangian displacements, respectively. We see that the
perturbative corrections to $\lan\bfL^{(1)2}\ran^{1/2}$ grow
proportionally to $t^{5/3}$ in the Einstein-de Sitter universe, in
agreement with Peebles (1969). The expressions between square brackets
give the generalisations of the results for a generic Friedmann
universe. Interestingly, all the corrections we have analysed are
proportional to the same invariant of the inertia tensor $\calJ$ of
the matter contained in the homogeneous Lagrangian volume $\Gamma$, a
result we can express as
\begin{itemize}
\item $\lan\bfL^{(1)2}\ran 
\propto \lan\bfL^{(2)2}\ran \propto 
\lan\bfL^{(1)}\!\cdot\bfL^{(3)}\ran 
\propto \mu_1^2 - 3\mu_2\;$,
\end{itemize}
\indent where $\mu_1$ and $\mu_2$ are the first and the second
invariant of the inertia tensor (see equations~(40) and (41)). This
invariant $\mu_1^2 - 3\mu_2$ has been thoroughly investigated in
Catelan \& Theuns (1996). As a consequence of this factorisation we
have been able to express the order of magnitude of the non-linear
corrections to $\lan\bfL^2\ran$ in terms of the linear contribution,
$\lan\bfL^2\ran = (1+\Upsilon)\,\lan\bfL^{(1)2}\ran$ (equation~(54)),
where $\Upsilon\approx 0.6$ for the standard CDM spectrum at galactic
scales. Taking into account that the non-linear correction is small,
we conclude that linear theory gives a good description of the angular
momentum up to maximum expansion. Since in addition linear theory
predicts, in the Einstein-de Sitter model, a growth rate $\bfL \propto
t$, it follows that the {\it initial} torque is a good estimate for
the tidal torque over the whole period during which the object is
spun up: $d\bfL(t)/dt\approx d\bfL(0)/dt$.

Finally, as is the case with almost any analytic calculation,
comparison of these results against observations is hampered by the
fact that the very final stages of galaxy formation are likely to be
highly non-linear and in addition dissipative processes may play an
important role as well. Analytic investigations are not able to take
such highly complex phenomena into account.

\section*{Acknowledgements}
We want to thank James Binney, George Efstathiou and Sabino Matarrese
for reading the original manuscript of this work. The Referee, Alan
Heavens, helped improve the presentation of the paper. PC and TT were
supported by the EEC Human Capital and Mobility Programme under
contracts CT930328 and CT941463 respectively.

{}

\section*{Appendix A}
In this first appendix we report the expressions of the Lagrangian
perturbative solutions $\bfS_n$ explicitly up to third-order and valid
for a generic non-flat universe. Specifically, we give the expressions
for the growing modes $D, E, F_a, F_b$ and $F_c$ for linear, second-
and third-order terms and the expressions for their wave vector
dependence, i.e. the Fourier transforms of the longitudinal potentials
$\psi^{(2)}, \psi_a^{(3)}, \psi_b^{(3)}$ and of the transverse
components $T_\al$. For brevity we use the symbol $\Theta(\tau)\equiv
{\rm ln}\sqrt{(\tau-1)/(\tau+1)}$ for the open universe case $(k=-1)$
and $\Lam(\tau) \equiv {\rm arctang}(1/\tau)=-i\Theta(i\tau)$ for the
closed universe case $(k=+1)$.\\\\

\noindent {\it A.1~ Zel'dovich approximation}\\

\noindent The growing mode of the density fluctuations $D(\tau)$ is given by
(Shandarin 1980):
\begin{equation}
D(\tau)=\f{5}{2}\left\{1+3\,(\tau^2-1)\left[1+\tau\,\Theta(\tau)\,\right]\right\}\;,
\end{equation}
for the open universe, and
\begin{equation}
D(\tau)=\f{5}{2}\left\{-1+3\,(\tau^2+1)\left[1-\tau\,\Lam(\tau)\,\right]
\right\}\;,
\end{equation}
for the closed universe. The solution~(61) can be obtained from~(60)
by substituting in the latter $\tau$ by $i\tau$ and reversing the sign
to make the growing mode positive. Note that, in contrast to Bouchet
\etal (1992) and Catelan (1995), we normalised $D(\tau)$ according to
the suggestion of Shandarin (1980): the coefficient $5/2$ is such that
$D(\tau)\rightarrow \tau^{-2}$ in the limit $\tau\rightarrow -\infty$,
which coincides with the Einstein-de Sitter case. $D(\tau)$ for the
different universes is plotted in Fig.~4.\\\\

\noindent{\it A.2~ Second-order approximation}\\

\noindent The time dependence of second-order growing mode $E(\tau)$
corresponding to the normalisation chosen for $D$ is:
\begin{equation}
E(\tau)= - \f{25}{8} -
\f{225}{8}\,(\tau^2-1)\left\{1+\tau\,\Theta(\tau) +\f{1}{2}\left[\tau
+(\tau^2-1)\,\Theta(\tau)\,\right]^2 \right\}\;,
\end{equation}
for the open universe and
\begin{equation}
E(\tau)= - \f{25}{8} +
\f{225}{8}\,(\tau^2+1)\left\{1-\tau\,\Lam(\tau)
-\f{1}{2}\left[\tau -(\tau^2+1)\,\Lam(\tau)\,\right]^2
\right\}\;,
\end{equation}
for the closed universe. These solutions have been first derived by
Bouchet \etal (1992). The extra factor $25/4$ of the present version
is due to the different normalisation of the first-order solution
$D$. An excellent approximation of the second-order growing mode is
$E=-\f{3}{7}D^2$ (see Fig.~4 for a plot of $E(\tau)$ and Fig.~5 for a
plot of the approximation). In the limit $\tau\rightarrow -\infty$ one
has $E=-\f{3}{7}\tau^{-4}$ which corresponds to the flat case.

The Fourier transform of the second-order potential $\psi^{(2)}$ is
(Catelan 1995):
\begin{equation}
\fpsi^{(2)}(\bfp) 
=-\f{1}{p^2}\int \f{d\bfp_1 d\bfp_2}{(2\pi)^6} 
\Big[(2\pi)^3 \de_D(\bfp_1+\bfp_2 - \bfp)\Big]\,
\kappa^{(2)}(\bfp_1,\bfp_2)
\,\fpsi^{(1)}(\bfp_1)\,\fpsi^{(1)}(\bfp_2)\;,
\end{equation}
where we have defined the kernel
\begin{equation}
\kappa^{(2)}(\bfp_1,\bfp_2)\equiv 
\f{1}{2}\,\Big[\,p_1^2\,p_2^2 - (\bfp_1\!\cdot\bfp_2)^2\Big] \;.
\end{equation}
This kernel describes the effects of non-linearity on the second-order
Lagrangian motion of the mass fluid elements.\\\\

\noindent{\it A.3~ Third-order approximation}\\

\noindent The third-order solution $\bfS_3$ actually corresponds to three
separable modes, two longitudinal and one transverse, as was discussed
previously (section 2.2). The expressions of the growing modes $F_a,
F_b$ and $F_c$ for a non-flat universe are known up to quadratures in
terms of the lower-order solutions $D$ and $E$. Explicitly (Catelan
1995):
\begin{equation}
F_a(\tau) = -2\,D(\tau)\int_{-\infty}^{\tau}d\tau_1\, D(\tau_1)^{-2}
\int_{-\infty}^{\tau_1} d\tau_2\, \al(\tau_2)\,D(\tau_2)^4\;,
\end{equation}
\begin{equation}
F_b(\tau) = -2\,D(\tau)\int_{-\infty}^{\tau}d\tau_1\, D(\tau_1)^{-2}
\int_{-\infty}^{\tau_1} d\tau_2 \,\al(\tau_2)\,D(\tau_2)^2
\left[E(\tau_2) - D(\tau_2)^2 \right]\;,
\end{equation}
\begin{equation}
F_c(\tau) = -\int_{-\infty}^{\tau}d\tau_1\int_{-\infty}^{\tau_1}d\tau_2\,\al(\tau_2)
\,D(\tau_2)^3\;.
\end{equation}
Excellent fits of these functions are $F_a(\tau)=-\f{1}{3}D^3,
F_b(\tau)=\f{10}{21}D^3$ and $F_c(\tau)=-\f{1}{7}D^3$ (see again
Fig.~4 for a plot of the functions and Fig.~5 for a plot of the
approximations; these functions were not shown in Catelan 1995). In
the limit $\tau\rightarrow -\infty$ one recovers the flat solutions as
required: $F_a(\tau)=-\f{1}{3}\tau^{-6},
F_b(\tau)=\f{10}{21}\tau^{-6}$ and $F_c(\tau)=-\f{1}{7}\tau^{-6}$.

The Fourier transforms of the third-order potentials $\psi_a^{(3)}$
and $\psi_b^{(3)}$ and of the transverse components $T_\al$ are
respectively (Catelan 1995)
\begin{equation}
\fpsi_a^{(3)}(\bfp) 
=-\f{1}{p^2}\int \f{d\bfp_1 d\bfp_2 d\bfp_3}{(2\pi)^9}\, 
[(2\pi)^3\de_D(\bfp_1+\bfp_2+\bfp_3 - \bfp)]\,
\kappa_a^{(3)}(\bfp_1,\bfp_2, \bfp_3)
\,\fpsi^{(1)}(\bfp_1)\,\fpsi^{(1)}(\bfp_2)\,\fpsi^{(1)}(\bfp_3),
\end{equation}
\begin{equation}
\fpsi_b^{(3)}(\bfp) 
=\f{1}{p^2}\int \f{d\bfp_1 d\bfp_2 d\bfp_3}{(2\pi)^9}\,
[(2\pi)^3\de_D(\bfp_1+\bfp_2+\bfp_3 - \bfp)]\,
\kappa_b^{(3)}(\bfp_1,\bfp_2, \bfp_3)
\,\fpsi^{(1)}(\bfp_1)\,\fpsi^{(1)}(\bfp_2)\,\fpsi^{(1)}(\bfp_3),
\end{equation}
\begin{equation}
\fT_\al^{(3)}(\bfp)=i\int \f{d\bfp_1 d\bfp_2 d\bfp_3}{(2\pi)^9}\,
[(2\pi)^3\de_D(\bfp_1+\bfp_2+\bfp_3 - \bfp)]\,
\iota_\al^{(3)}(\bfp_1,\bfp_2, \bfp_3)
\,\fpsi^{(1)}(\bfp_1)\,\fpsi^{(1)}(\bfp_2)\,\fpsi^{(1)}(\bfp_3),
\end{equation}
where we have introduced the following kernels:
\begin{equation}
\kappa_a^{(3)}(\bfp_1,\bfp_2, \bfp_3)\equiv 
\f{1}{6}\,\epsilon_{\al\gamma\de}\,\epsilon_{\beta\eta\s}\,
p_{\al}\,p_{1\beta}\,p_{2\gamma}\,p_{2\eta}\,p_{3\de}\,p_{3\s},
\end{equation}
\begin{equation}
\kappa_b^{(3)}(\bfp_1,\bfp_2, \bfp_3)\equiv
\f{1}{2}\left[
\bfp\cdot\bfp_1-\f{\bfp_1\!\cdot(\bfp_2+\bfp_3)}{|\bfp_2+\bfp_3|}
\f{\bfp\cdot(\bfp_2+\bfp_3)}{|\bfp_2+\bfp_3|}
\right]\,\kappa^{(2)}(\bfp_2,\bfp_3)\;,
\end{equation}
\begin{equation}
\iota_\al^{(3)}(\bfp_1,\bfp_2, \bfp_3)\equiv
\f{1}{2}\,\f{\bfp_1\!\cdot(\bfp_2+\bfp_3)}{|\bfp_2+\bfp_3|^2}\,
\kappa^{(2)}(\bfp_2,\bfp_3)\,(\bfp_2+\bfp_3-\bfp_1)_\al\;.
\end{equation}
These expressions are more suitable to compute correction terms to
$\bfL$ than the original equations as derived in Catelan (1995).

\begin{figure}
\setlength{\unitlength}{1cm}
\centering
\begin{picture}(10,12)
\put(0.0,0.0){\includegraphics{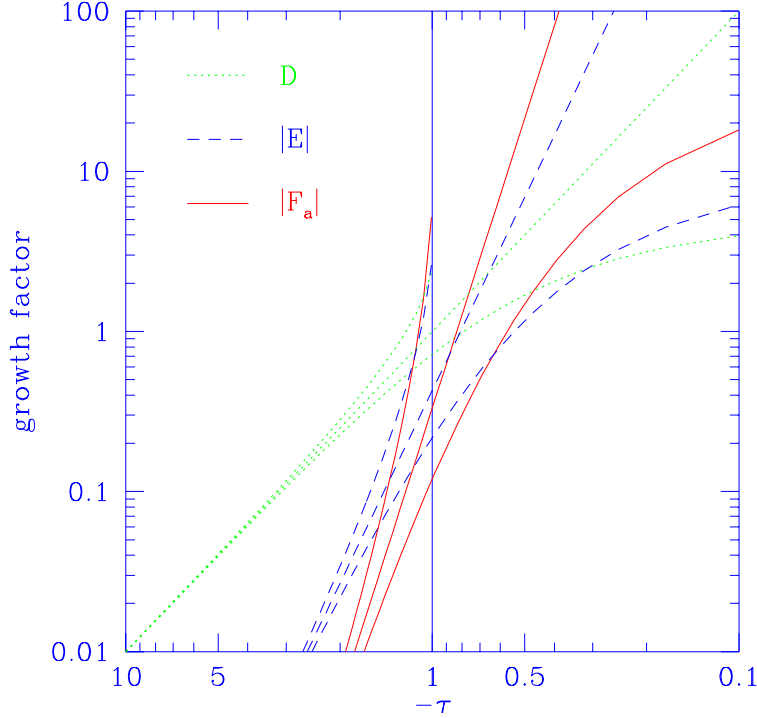}}
\end{picture}
\caption{Growth factors of the first, second and third-order
Lagrangian displacements, respectively $D(\tau)$, $E(\tau)$ and
$F_a(\tau)$, for open (upper curves), flat (middle curves) and closed
(lower curves) universes. The remaining third-order growth factors
$F_b(\tau)$ and $F_c(\tau)$ are not shown as they are almost identical
to $F_a(\tau)$.}
\label{fig:Shan}
\end{figure}

\begin{figure}
\setlength{\unitlength}{1cm}
\centering
\begin{picture}(10,12)
\put(0.0,0.0){\includegraphics{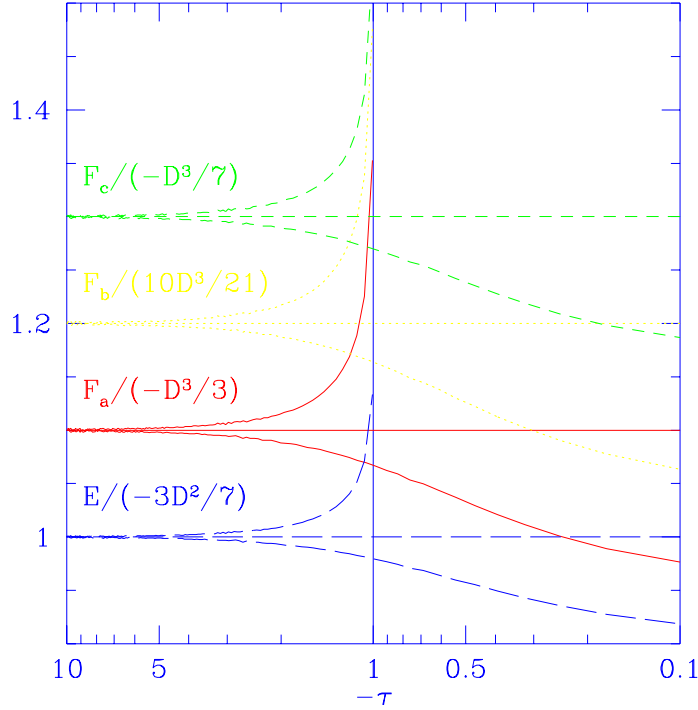}}
\end{picture}
\caption{Ratios between the higher-order growth factors $E$, $F_a$,
$F_b$ and $F_c$ and their approximations in terms of powers of $D$,
for open (upper curves), flat (middle curves) and closed (lower
curves) universes. The third-order ratios have been offset
artificially by 0.1, 0.2 and 0.3 for clarity. The second-order ratio
$E/(-3D^2/7)$ is shown here for completeness; it has been first shown
as a function of $\Omega$ in Bouchet et al. (1992).}
\label{fig:ratios}
\end{figure}

\section*{Appendix B}
In this appendix we give explicit derivations for the ensemble
averages $\lan\bfL^{(2)2}\ran$,
$\lan\bfL^{(1)}\!\cdot\bfL_h^{(3)}\ran$ and
$\lan\bfL^{(1)}\!\cdot\bfL^{(12)}\ran$.  We will use the perturbative
corrections to the Lagrangian displacement as reviewed previously.\\

\noindent $\bullet$ Starting with the former one, we use the expression~(26) for
$L^{(2)}_\al$ to find:
\begin{equation}
\lan\bfL^{(2)2}\ran=
\dot{E}(\tau)^2\,\eps_{\al\beta\gamma}\,\eps_{\al\beta'\gamma'}\,
\calJ_{\s\gamma}\,\calJ_{\s'\gamma'}\,
\lan\,\calD^{(2)}_{\beta\s}\calD^{(2)}_{\beta'\s'}\ran\;.
\end{equation}
The second-order deformation tensor $\calD^{(2)}_{\al\beta}$ may be
written in terms of the second-order potential $\psi^{(2)}$ as
\begin{equation}
\calD^{(2)}_{\al\beta}\equiv \p_\al\,\p_\beta \psi^{(2)}({\bf 0})=
-\int\f{d\bfp}{(2\pi)^3}\,p_{\al}\,p_{\beta}\,\fpsi^{(2)}(\bfp)\,\fW(pR)\;,
\end{equation}
where the field $\psi^{(2)}$ is now assumed to be filtered on scales
$R$ using the smoothing function $W_R$. Inserting the expression~(64)
for the Fourier transform $\fpsi^{(2)}(\bfp)$ of $\psi^{(2)}$ we find
\begin{equation}
\calD^{(2)}_{\al\beta}=
\int\f{d\bfp_1\,d\bfp_2}{(2\pi)^6}\,
\f{(\bfp_1+\bfp_2)_\al\,(\bfp_1+\bfp_2)_\beta}{|\bfp_1+\bfp_2|^2}\,
\fW(|\bfp_1+\bfp_2|R)\,\kappa^{(2)}(\bfp_1, \bfp_2)\,
\fpsi^{(1)}(\bfp_1)\,\fpsi^{(1)}(\bfp_2)\;.
\end{equation}
From this last expression we obtain:
\begin{eqnarray}
\lan\,\calD^{(2)}_{\beta\s}\calD^{(2)}_{\beta'\s'}\ran&=&
\int \f{d\bfp_1\,d\bfp_2}{(2\pi)^6}\,\f{d\bfp_1'\,d\bfp_2'}{(2\pi)^6}\,
\fW(|\bfp_1+\bfp_2|R)\,\fW(|\bfp_1'+\bfp_2'|R)\,
\kappa^{(2)}(\bfp_1, \bfp_2)\,\kappa^{(2)}(\bfp_1', \bfp_2')\,
\nonumber \\
&\times&
\f{(\bfp_1+\bfp_2)_\beta\,(\bfp_1+\bfp_2)_\s}{|\bfp_1+\bfp_2|^2}\,
\f{(\bfp_1'+\bfp_2')_{\beta'}\,(\bfp_1'+\bfp_2')_{\s'}}{|\bfp_1'+\bfp_2'|^2}\,
\lan\,\fpsi^{(1)}(\bfp_1)\,\fpsi^{(1)}(\bfp_2)\,
\fpsi^{(1)}(\bfp_1')\,\fpsi^{(1)}(\bfp_2') \,\ran\;.
\end{eqnarray}
Since the primordial gravitational potential is assumed to be Gaussian
distributed, one has (see, e.g., Peebles 1980)
\begin{eqnarray}
\lan\,\fpsi^{(1)}(\bfp_1)\,\fpsi^{(1)}(\bfp_2)\,
\fpsi^{(1)}(\bfp_1')\,\fpsi^{(1)}(\bfp_2') \,\ran &=&
(2\pi)^3\,\de_D(\bfp_1 +\bfp_2)\,P_\psi(p_1)\,
(2\pi)^3\,\de_D(\bfp_1' +\bfp_2')\,P_\psi(p_1')
\nonumber \\
&+&
(2\pi)^3\,\de_D(\bfp_1 +\bfp_1')\,P_\psi(p_1)\,
(2\pi)^3\,\de_D(\bfp_2 +\bfp_2')\,P_\psi(p_2)
\nonumber \\
&+&
(2\pi)^3\,\de_D(\bfp_1 +\bfp_2')\,P_\psi(p_1)\,
(2\pi)^3\,\de_D(\bfp_2 +\bfp_1')\,P_\psi(p_2)\;.
\end{eqnarray}
The first term does not contribute to the integral since
$\kappa^{(2)}(\bfp, -\bfp)=0$ . The remaining two terms give
identical contributions, hence:
\begin{equation}
\lan\calD^{(2)}_{\beta\s}\calD^{(2)}_{\beta'\s'}\ran=
2\!\int\!\!\f{d\bfp_1\,d\bfp_2}{(2\pi)^6}\,
\f{(\bfp_1+\bfp_2)_\beta\,(\bfp_1+\bfp_2)_\s\,
(\bfp_1+\bfp_2)_{\beta'}\,(\bfp_1+\bfp_2)_{\s'}}
{|\bfp_1+\bfp_2|^4}\,
[\kappa^{(2)}(\bfp_1, \bfp_2)]^2\,
[\fW(|\bfp_1+\bfp_2|R)]^2 P_\psi(p_1) P_\psi(p_2).
\end{equation}
The trick now is to reduce this integral in such a way that we can apply
the rule,
\begin{equation}
\int_{sphere}d\bfp\,p_{\al}\,p_{\beta}\,p_{\gamma}\,p_{\de}\,F(|\bfp|)=
\f{4\pi}{15}
(\de_{\al\beta}\,\de_{\gamma\de}+\de_{\al\gamma}\,\de_{\beta\de}+
\de_{\al\,\de}\,\de_{\beta\,\gamma})\int dp\,p^6\,F(p)\;,
\end{equation}
which holds for any function $F(p)$ which depends only on the modulus
$p$ of $\bfp$. Let us define the new variables $\bfk_1\equiv
\bfp_1+\bfp_2$ and $\bfk_2\equiv \bfp_1-\bfp_2$. The determinant of
the Jacobian of this transformation is $1/8$. At this point, noting
that
\begin{equation}
p_1^{-4}\,p_2^{-4}\,[\kappa^{(2)}(\bfp_1, \bfp_2)]^2 =
4\left[\f{1-\theta^2}
{(k_1\,k_2^{-1}+k_2\,k_1^{-1})^2-4\theta^2}\right]^2\;, \nonumber
\end{equation}
where $\theta\equiv \bfk_1\!\cdot\bfk_2/k_1\,k_2$ and $k_h\equiv|\bfk_h|$,
the integral~(80) may be written as
\be
\lan\calD^{(2)}_{\beta\s}\calD^{(2)}_{\beta'\s'}\ran=
\int\!\!\f{d\bfk_1\,d\bfk_2}{(2\pi)^6}\,
\f{k_{1\beta}\,k_{1\s}\,k_{1\beta'}\,k_{1\s'}}
{k_1^4}\,
\left[\f{1-\theta^2}
{(k_1\,k_2^{-1}+k_2\,k_1^{-1})^2-4\theta^2}\right]^2\,
[\fW(k_1\,R)]^2\, P_\delta(|\bfk_1+\bfk_2|/2)\, P_\delta(|\bfk_1-\bfk_2|/2)\;.
\end{equation}
This expression holds for any power spectrum
$P_\delta(p)=p^4\,P_\psi(p)$, but let us assume for simplicity that
the power spectrum is scale-free, $P_\psi(p) = Ap^{n-4}$. The
calculation can be continued as follows. Since one has:
\begin{equation}
|\bfk_1+\bfk_2|^n\,|\bfk_1-\bfk_2|^n=
(k_1\,k_2)^n\,
\big[(k_1\,k_2^{-1}+k_2\,k_1^{-1})^2-4\theta^2\big]^{n/2}\;,
\nonumber
\end{equation}
the integral in~(83) may be simplified to
\begin{equation}
\lan\calD^{(2)}_{\beta\s}\calD^{(2)}_{\beta'\s'}\ran \!=\!
\f{A^2}{4^n}\!
\int\!\! \f{d\bfk_1}{(2\pi)^3}\,
\f{k_{1\beta}\,k_{1\s}\,k_{1\beta'}\,k_{1\s'}}
{k_1^{4-n}}\,
[\fW(k_1\,R)]^2
\int_0^\infty \!\f{d k_2}{(2\pi)^2}\,k_2^{n+2}
\int_{-1}^{+1}\!
d\theta\, \left(1-\theta^2\right)^2 \,
\left[\Big(\f{k_1}{k_2}+\f{k_2}{k_1}\Big)^2-4\theta^2\right]^{(n-4)/2}\;.
\end{equation}
We note at this point that the function
\begin{equation}
2^{2n-1}\,\Delta^{(2)}(k_1; n)\equiv 
\int_0^\infty dk_2\,k_2^{n+2}\int_{-1}^{+1}
d\theta\, \left(1-\theta^2\right)^2 \,
\big[(k_1\,k_2^{-1}+k_2\,k_1^{-1})^2-4\theta^2\big]^{(n-4)/2}\;,
\end{equation}
which typically has to be evaluated numerically, depends only on the
modulus $|\bfk_1|$ and on the spectral index $n$, as indicated. We can
therefore apply the rule in (81) since the smoothing function depends
only on the modulus of $\bfk_1$ as well, to do the transformation:
\begin{eqnarray}
\lan\calD^{(2)}_{\beta\s}\calD^{(2)}_{\beta'\s'}\ran&=&
\f{A^2}{2(2\pi)^2}
\int \f{d\bfk}{(2\pi)^3}\,
k_{\beta}\,k_{\s}\,k_{\beta'}\,k_{\s'}\,
[\fW(kR)]^2\,
k^{n-4}\,
\Delta^{(2)}(p_1; n)
\nonumber \\
&=&
\f{1}{15}\,
(\de_{\beta\s}\,\de_{\beta'\s'}+\de_{\beta\beta'}\,\de_{\s\s'}+
\de_{\beta\s'}\,\de_{\beta'\s})\,\Sigma^{(2)}(R; n)\;,
\end{eqnarray}
where $\Sigma^{(2)}$ is given in equation~(43). Finally, following
along the lines of the the derivation of $\lan L^{(1)2}\ran$
reported in Appendix A of Catelan and Theuns (1996), one ends up with
the result~(42) in the main text.

The last expressions are not valid for more physical, non-power law,
spectra. However, the appropriate expressions can be found by
following the same strategy. Let us consider for example the case of
the CDM power spectrum (see, e.g., Efstathiou 1989), where the power
spectrum is parametrised by:
\begin{equation}
P_\psi(p)\equiv A\,p^{-3}\,[T(p)]^2 = A\,p^{-3}\,
\left[1+\Big(ap +(bp)^{3/2}+ (cp)^2\Big)^\nu\right]^{-2/\nu}\;,
\end{equation}
where $a=6.4\,(\Om h^2)^{-1}$ Mpc,~$b=3.0\,(\Om h^2)^{-1}$ Mpc$,~
c=1.7\,(\Om h^2)^{-1}$ Mpc, and $\nu=1.13$; $A$ is the normalisation
of the spectrum, as before. In this case, one obtains instead of
equation~(85):
\begin{eqnarray}
\lan\calD^{(2)}_{\beta\s}\calD^{(2)}_{\beta'\s'}\ran &=&
\f{A^2}{4}
\int \f{d\bfk_1}{(2\pi)^3}\,
\f{k_{1\beta}\,k_{1\s}\,k_{1\beta'}\,k_{1\s'}}
{k_1^{3}}\,
[\fW(k_1\,R)]^2
 \nonumber \\
&\times&\int_0^\infty \!\f{d k_2}{(2\pi)^2}\,k_2^{3}
\int_{-1}^{+1}\!
d\theta\, \left(1-\theta^2\right)^2 \,
\big[(k_1\,k_2^{-1}+k_2\,k_1^{-1})^2-4\theta^2\big]^{-3/2}
\,[\calT(k_1, k_2; +\theta)]^2\,[\calT(k_1, k_2; -\theta)]^2\;,
\end{eqnarray}
where we have defined the transfer functions,
\begin{equation}
\calT(k_1, k_2; \pm\theta)^2 \equiv
\left\{1+
\left[a\,G(\pm\theta) +
\Big(b\,G(\pm\theta)\Big)^{3/2}+ 
\Big(c\,G(\pm\theta)\Big)^2\right]^\nu\right\}^{-2/\nu}\;,
\end{equation}
and $G(\pm\theta)\equiv
\f{1}{2}\sqrt{k_1k_2\,}\,(k_1\,k_2^{-1}+k_2\,k_1^{-1}\pm
2\theta)^{1/2}$.  As a consequence of the change of variables,
$(\bfp_1,\bfp_2)\rightarrow (\bfk_1,\bfk_2)$ we have, for example,
$\calT(k_1, k_2; \theta)= T(p_1)$. Finally, proceeding as earlier we get:
\begin{equation}
\Sigma_{CDM}^{(2)}(R) \equiv\f{A^2}{(2\pi)^4} \int_0^\infty dp
\,p^{3}\,[\fW(pR)]^2\, \Delta_{CDM}^{(2)}(p)\;,
\end{equation}
\begin{equation}
\Delta_{CDM}^{(2)}(k_1)\equiv 
\f{1}{2}\int_0^\infty dk_2\,k_2^{3}\int_{-1}^{+1}
d\theta\, \left(1-\theta^2\right)^2 \,
\big[(k_1\,k_2^{-1}+k_2\,k_1^{-1})^2-4\theta^2\big]^{-3/2}
\,[\calT(k_1, k_2; +\theta)]^2\,[\calT(k_1, k_2; -\theta)]^2\;.
\end{equation}
~\\

\noindent$\bullet$ Let us now summarise how to deal with the simpler
case of the averages $\lan\bfL^{(1)}\!\cdot\bfL_h^{(3)}\ran$.  We first
show that $\lan\,\bfL^{(1)}\!\cdot\bfL^{(3)}_a\ran=0$. One has:
\begin{equation}
\lan\bfL^{(1)}\!\cdot\bfL_a^{(3)}\ran=
\sum_{\al}\lan L^{(1)}_{\al} L^{(3)}_{a\,\al}\ran=
\dot{D}(\tau)\,\dot{F_a}(\tau)\,\eps_{\al\beta\gamma}\,\eps_{\al\beta'\gamma'}\,
\calJ_{\s\gamma}\,\calJ_{\s'\gamma'}\,
\lan\,\calD^{(1)}_{\beta\s}\calD^{(3)}_{a\,\beta'\s'}\ran\;.
\end{equation}
Writing $\calD^{(3)}_{a\,\beta'\s'}$ as
\begin{equation}
\calD^{(3)}_{a\,\beta'\s'}\equiv \p_{\beta'}\,\p_{\s'} \psi^{(3)}_a({\bf 0})=
-\int\f{d\bfp}{(2\pi)^3}\,p_{\beta'}\,p_{\s'}\,\fpsi^{(3)}_a(\bfp)\,\fW(pR)\;,
\end{equation}
and inserting the expression~(64) for $\fpsi^{(3)}_a(\bfp)$, one gets
\begin{eqnarray}
\lan\,\calD^{(1)}_{\beta\s}\calD^{(3)}_{a\,\beta'\s'}\ran
&=&
\f{1}{6}\,\epsilon_{\al''\gamma''\de''}\,\epsilon_{\beta''\eta''\s''}\,
\int \f{d\bfp_1 d\bfp_2}{(2\pi)^6}\,p_2^{-2}\,
[\fW(p_2R)]^2\,P_\psi(p_1)\,P_\psi(p_2)\,
p_{2\beta'}\,p_{2\s'}\,p_{2\al''}\,p_{2\beta}\,p_{2\s}
\nonumber \\
&\times&
\Big[p_{1\beta''}\,p_{1\gamma''}\,p_{1\eta''}\,p_{2\delta''}\,p_{2\s''}+
\,p_{1\beta''}\,p_{2\gamma''}\,p_{2\eta''}\,p_{1\delta''}\,p_{1\s''}+
p_{2\beta''}\,p_{1\gamma''}\,p_{1\eta''}\,p_{1\delta''}\,p_{1\s''}\Big]\;,
\end{eqnarray}
which is zero for any $W$ and power spectrum $P_\psi$, since symmetric
tensors saturate antisymmetric tensors. This completes the proof.\\

\noindent$\bullet$ Next, let us compute:
\begin{equation}
\lan\bfL^{(1)}\!\cdot\bfL_b^{(3)}\ran=
\sum_{\al}\lan L^{(1)}_{\al} L^{(3)}_{b\,\al}\ran=
\dot{D}(\tau)\,\dot{F_b}(\tau)\,\eps_{\al\beta\gamma}\,\eps_{\al\beta'\gamma'}\,
\calJ_{\s\gamma}\,\calJ_{\s'\gamma'}\,
\lan\,\calD^{(1)}_{\beta\s}\calD^{(3)}_{b\,\beta'\s'}\ran\;.
\end{equation}
The third-order deformation tensor $\calD^{(3)}_{b\,\al\beta}$
may be written as
\begin{equation}
\calD^{(3)}_{b\,\al\beta}\equiv \p_\al\,\p_\beta \psi^{(3)}_b({\bf 0})=
-\int\f{d\bfp}{(2\pi)^3}\,p_{\al}\,p_{\beta}\,\fpsi^{(3)}_b(\bfp)\,\fW(pR)\;,
\end{equation}
where the field $\psi^{(3)}_b$ is now filtered on scale $R$ as
previously. Inserting the expression~(65) for $\fpsi^{(3)}_b(\bfp)$,
we get:
\begin{eqnarray}
\calD^{(3)}_{b\,\al\beta}=&-&\int\f{d\bfp_1\,d\bfp_2\,d\bfp_3}{(2\pi)^9}\,
\f{(\bfp_1+\bfp_2+\bfp_3)_\al\,(\bfp_1+\bfp_2+\bfp_3)_\beta}
{|\bfp_1+\bfp_2+\bfp_3|^2}\,
\fW(|\bfp_1+\bfp_2+\bfp_3|R) \nonumber \\
&\times&
\f{1}{2}\,p^2_1\,
\left[
1-\Big(\f{\bfp_1\!\cdot(\bfp_2+\bfp_3)}{p_1|\bfp_2+\bfp_3|}\Big)^2 \right]
\,\kappa^{(2)}(\bfp_2, \bfp_3)\,
\fpsi^{(1)}(\bfp_1)\,\fpsi^{(1)}(\bfp_2)\,\fpsi^{(1)}(\bfp_3)\;.
\end{eqnarray}
The average can now be computed easily by following the same steps as
in the derivation of $\lan L^{(2)2}\ran$. One ends up with the
expression
\begin{equation}
\lan\,\calD^{(1)}_{\beta\s}\calD^{(3)}_{b\,\beta'\s'}\ran=
\int\!\!\f{d\bfp_2}{(2\pi)^3}\,
p_{2\beta}\,p_{2\s}\,p_{2\beta'}\,p_{2\s'}\,
[\fW(p_2R)]^2\,p_2\,P_\psi(p_2)\,
\int_0^\infty\!\f{dp_1}{(2\pi)^2}\,p_1^5\,P_\psi(p_1)
\int_{-1}^{+1}d\theta\;
\f{(1-\theta^2)^2}
{p_1\,p_2^{-1}+p_2\,p_1^{-1}-2\theta}\;.
\end{equation}
We stress that this results holds for any spectrum $P_\psi(p)$,
including the CDM spectrum~(88). Limiting ourselves to a scale-free
power spectrum $P_\psi(p)=Ap^n$ and applying the rule (81), one
recovers the equation~(46) in the main text. The generalisation to a
CDM power spectrum is straightforward, since there is no coupling
between different wave vectors in the power spectrum in this case.\\

\noindent$\bullet$ In a similar fashion, one computes the average
\begin{equation}
\lan\bfL^{(1)}\!\cdot\bfL_c^{(3)}\ran=
\sum_{\al}\lan L^{(1)}_{\al} L^{(3)}_{c\,\al}\ran=
\dot{D}(\tau)\,\dot{F_c}(\tau)\,\eps_{\al\beta\gamma}\,\eps_{\al\beta'\gamma'}\,
\calJ_{\s\gamma}\,\calJ_{\s'\gamma'}\,
\lan\,\calD^{(1)}_{\beta\s}\,T^{(3)}_{\gamma'\s'}\ran\;.
\end{equation}
From the expression~(71) for $\fT^{(3)}_\al(\bfp)$ we have:
\begin{eqnarray}
T_{\al\beta}^{(3)}&=&i\int\f{d\bfp}{(2\pi)^3}\,p_{\beta}\,\,
\fT^{(3)}_\al(\bfp)\,\fW(pR)
\nonumber \\
&=&-\int\f{d\bfp_1\,d\bfp_2\,d\bfp_3}{(2\pi)^9}\,
(\bfp_1+\bfp_2+\bfp_3)_\beta\;
\iota^{(3)}_\al(\bfp_1, \bfp_2, \bfp_3)\,
\fW(|\bfp_1+\bfp_2+\bfp_3|R)\, 
\fpsi^{(1)}(\bfp_1)\,\fpsi^{(1)}(\bfp_2)\,\fpsi^{(1)}(\bfp_3)\;.
\end{eqnarray}
One now proceeds as earlier. For any power spectrum $P_\psi(p)$, the
result is:
\begin{equation}
\lan\,\calD^{(1)}_{\beta\s}\,T^{(3)}_{\gamma'\s'}\ran=
\int\!\!\f{d\bfp_2}{(2\pi)^3}\,
p_{2\beta}\,p_{2\s}\,p_{2\beta'}\,p_{2\s'}\,
[\fW(p_2R)]^2\,p_2^2\,P_\psi(p_2)\,
\int_0^\infty\!\f{dp_1}{(2\pi)^2}\,p_1^4\,P_\psi(p_1)
\int_{-1}^{+1}d\theta\;
\f{(1-\theta^2)\,(\theta-p_1\,p_2^{-1})}
{p_1\,p_2^{-1}+p_2\,p_1^{-1}-2\theta}\;.
\end{equation}
Assuming a scale-free power spectrum $P_\psi(p)=Ap^n$ and applying
again the rule~(81), one recovers the equation~(47) in the main
text. Once again, the generalisation to a CDM power spectrum is
straightforward.\\

\noindent$\bullet$ We conclude this appendix by discussing the
third-order correction $\bfL^{(12)}$:
\begin{equation}
\bfL^{(12)}(\tau)=\eta_0\,(D\dot{E}-E\dot{D})\!\int_\Gamma d\bfq\,
\nabla\psi^{(1)}\times\nabla\psi^{(2)}\;.
\end{equation}
Expanding the potentials in Taylor series around $\bfq={\bf 0}$, one
obtains for the $\al$-component 
\begin{equation}
L^{(12)}_\al=(D\dot{E}-E\dot{D})\,\eps_{\al\beta\gamma}
\Big[\eta_0\,\Gamma\,\p_\beta\psi^{(1)}({\bf 0})\,\p_\gamma\psi^{(2)}({\bf 0})
+\calD^{(1)}_{\beta\s}\,\calD^{(2)}_{\gamma\eta}\,\calJ_{\s\eta}\Big]
\equiv L^{(12)}_{A\,\al}+L^{(12)}_{B\,\al}\;.
\end{equation}
In general, the first of these two terms may be written as
\begin{eqnarray}
L^{(12)}_{A\,\al} &=&
\eta_0\,\Gamma\,(D\dot{E}-E\dot{D})\,\eps_{\al\beta\gamma}\,
\int\f{d\bfp_1\,d\bfp_2\,d\bfp_3}{(2\pi)^9}\,p_{1\beta}\,
\f{(\bfp_2+\bfp_3)_{\gamma}}
{|\bfp_2+\bfp_3|^2}\,\fW(p_1R)\,\fW(|\bfp_2+\bfp_3|R) \nonumber \\
&\times&
\,\kappa^{(2)}(\bfp_2, \bfp_3)\,
\fpsi^{(1)}(\bfp_1)\,\fpsi^{(1)}(\bfp_2)\,\fpsi^{(1)}(\bfp_3)\;,
\end{eqnarray}
and the second term as
\begin{eqnarray}
L^{(12)}_{B\,\al} &=&
-(D\dot{E}-E\dot{D})\,\eps_{\al\beta\gamma}\,\calJ_{\s\eta}
\int\f{d\bfp_1\,d\bfp_2\,d\bfp_3}{(2\pi)^9}\,p_{1\beta}\,p_{1\s}\,
\f{(\bfp_2+\bfp_3)_{\gamma}\,(\bfp_2+\bfp_3)_{\eta}}
{|\bfp_2+\bfp_3|^2}\,\fW(p_1R)\,\fW(|\bfp_2+\bfp_3|R) \nonumber \\
&\times&
\,\kappa^{(2)}(\bfp_2, \bfp_3)\,
\fpsi^{(1)}(\bfp_1)\,\fpsi^{(1)}(\bfp_2)\,\fpsi^{(1)}(\bfp_3)\;.
\end{eqnarray}
With these informations, one can compute the ensemble average
$\lan\bfL^{(1)}\!\cdot\bfL^{(12)}\ran=
\lan\bfL^{(1)}\!\cdot\bfL_A^{(12)}\ran+
\lan\bfL^{(1)}\!\cdot\bfL_B^{(12)}\ran$. The resulting expression are, 
respectively,
\begin{eqnarray}
\lan\bfL^{(1)}\!\cdot\bfL_A^{(12)}\ran &=&
-\eta_0\,\Gamma\,\dot{D}(D\dot{E}-E\dot{D})
\int\f{d\bfp_1\,d\bfp_2}{(2\pi)^6}\,\fW(p_1R)\,\fW(p_2R)\,
\fW(|\bfp_1-\bfp_2|R)\,P_\psi(p_1)\,P_\psi(p_2)
\nonumber \\
&\times&|\bfp_1-\bfp_2|^{-2}\,\kappa^{(2)}(\bfp_1, \bfp_2)\,
p_{2\al}(p_{1\al}p_{2\beta}-p_{2\al}p_{1\beta})
p_{2\gamma}\,\calJ_{\beta\gamma}\nonumber\\
&\equiv& -\eta_0\Gamma\,\calF\left[p_{2\al}(p_{1\al}p_{2\beta}-p_{2\al}p_{1\beta})
p_{2\gamma}\,\calJ_{\beta\gamma}\right]\;
\end{eqnarray}
\begin{eqnarray}
\lan\bfL^{(1)}\!\cdot\bfL_B^{(12)}\ran &=&
\dot{D}(D\dot{E}-E\dot{D})
\int\f{d\bfp_1\,d\bfp_2}{(2\pi)^6}\,\fW(p_1R)\,\fW(p_2R)\,
\fW(|\bfp_1-\bfp_2|R)\,P_\psi(p_1)\,P_\psi(p_2)
\nonumber \\
&\times&|\bfp_1-\bfp_2|^{-2}\,\kappa^{(2)}(\bfp_1, \bfp_2)\,
p_{2\al}(p_{1\al}p_{2\beta}-p_{2\al}p_{1\beta})
p_{2\gamma}\,\calJ_{\beta\gamma}\,
(p_{1\de}(\bfp_2-\bfp_1)_\eta\,\calJ_{\de\eta})\nonumber\\
&\equiv & \calF\left[p_{2\al}(p_{1\al}p_{2\beta}-p_{2\al}p_{1\beta})
p_{2\gamma}\,\calJ_{\beta\gamma}\,
(p_{1\de}(\bfp_2-\bfp_1)_\eta\,\calJ_{\de\eta})\right]\;,
\end{eqnarray}
where we have introduced the integral operator $\calF$ for
conciseness.  We proceed by showing that the first term
$\lan\bfL^{(1)}\!\cdot\bfL_A^{(12)}\ran$ is zero for any power
spectrum. In the eigenframe of the inertia tensor and taking advantage
of the fact that $\calF$ is a linear operator, we find:
\begin{equation}
\lan\bfL^{(1)}\!\cdot\bfL_A^{(12)}\ran = -\eta_0\Gamma\,
\iota_\al\,\calF[A_\al]
\end{equation}
where we defined
$A_\al\equiv(\bfp_1\cdot\bfp_2)p^2_{2\al}-\bfp^2_2p_{1\al}p_{2\al}$.
Using the fact that, because of rotational symmetry, $\calF[A_\al]=
\calF[A_\beta]$ for any Cartesian axes $\al\neq\beta$, the result
simplifies to
\begin{equation}
\lan\bfL^{(1)}\!\cdot\bfL_A^{(12)}\ran =
-\eta_0\Gamma\,\mu_1\,\calF[A_z]\;.
\end{equation}
Since $\Sigma_\al A_\al=0$ we finally get $\calF[\Sigma_\al A_\al]=0
={1\over 3}\calF[A_z]$, which completes the proof.

We can use a similar trick to simplify the second term: 
defining $B_\al\equiv p_{1\al}(\bfp_2-\bfp_1)_\al$ (no sum over
$\alpha$!) it is straightforward to show that in the
eigenframe of $\calJ$ one gets 
\begin{eqnarray}
\lan\bfL^{(1)}\!\cdot\bfL_B^{(12)}\ran
&=& \calF[\iota_\al A_\al\, \iota_\beta B_\beta]\nonumber\\
&=& \calF[ \iota^2_x A_x B_x +\iota^2_y A_y B_y +\iota^2_z A_z B_z
+ \iota_x \iota_y (A_x B_y + A_y B_x)+ \iota_x \iota_z (A_x B_z + A_z B_x)
+ \iota_y \iota_z (A_y B_z + A_z B_y)]\nonumber\\
&=&(\mu_1^2-2\mu_2) \calF[A_z B_z] + 2\mu_2 \calF[A_z B_x]\nonumber\\
&=&(\mu_1^2-3\mu_2) \calF[A_z B_z]\;.
\end{eqnarray}
The last equality is obtained as follows: we first use the same trick
as before to prove that $\calF[\Sigma_\al B_\al\, A_z] = {1\over 3}
\calF[\Sigma_\al B_\al\, \Sigma_\beta A_\beta] = 0$. Next, since because
of symmetry $\calF[A_z B_x]=\calF[A_z B_y]$ we get $\calF[A_z B_x]=
-{1\over 2} \calF[A_z B_z]$. The integral $\calF[A_z B_z]$ has been
computed numerically. As in all previously discussed correction terms,
$\mu_1^2-3\mu_2$ factors out. One recovers equation~(53).

\begin{thebibliography}{}

\bibitem[\protect\citename{Barnes \& Efstathiou } 1987] {be87}
Barnes J. \& Efstathiou G., 1987, ApJ, 319, 575

\bibitem[\protect\citename{Bouchet \etal} 1992] {bjcp92}
Bouchet F.R., Juszkiewicz R., Colombi S. \& Pellat R., 1992, ApJ, 394, L5

\bibitem[\protect\citename{Buchert} 1992] {b92}
Buchert T., 1992, MNRAS, 254, 729

\bibitem[\protect\citename{Buchert} 1994] {b94}
Buchert T., 1994, MNRAS, 267, 811

\bibitem[\protect\citename{Catelan} 1995] {c95}
Catelan P., 1995, MNRAS, 276, 115

\bibitem[\protect\citename{Catelan \& Theuns} 1996] {ct96}
Catelan P. \& Theuns T., 1996, MNRAS, to be published

\bibitem[\protect\citename{Doroshkevich } 1970] {dor70}
Doroshkevich A.G., 1970, Astrofizika, 6, 581

\bibitem[\protect\citename{Efstathiou} 1989] {e89}
Efstathiou G., 1989, in {\it Physics of the Early Universe}, 
eds. Peacock J.A., Heavens A.F. \& Davies A.T., Edinburgh University
Press

\bibitem[\protect\citename{Eisenstein \& Loeb} 1995] {el95}
Eisenstein D.J. \& Loeb A., 1995, ApJ, 439, 520

\bibitem [\protect\citename{Heavens \& Peacock }1988]{h88}
Heavens A. \& Peacock J., 1988, MNRAS, 232, 339

\bibitem [\protect\citename{Hoffman }1986]{h86}
Hoffman Y., 1986, ApJ, 301, 65

\bibitem [\protect\citename{Hoffman }1988]{h88}
Hoffman Y., 1988, ApJ, 329, 8

\bibitem [\protect\citename{Hoffman \& Ribak}1991]{hr91}
Hoffman Y. \& Ribak E., 1991, ApJ, 380, L5

\bibitem[\protect\citename{Hoyle } 1949] {h49}
Hoyle F., 1949, in {\it Problems of Cosmical Aerodynamics}, eds.
Burgers J.M. \& van de Hulst H.C. (Central Air Documents Office, Dayton)

\bibitem[\protect\citename{Kofman \& Pogosyan } 1995] {kp95}
Kofman L. \& Pogosyan D., 1995, ApJ, 442, 30

\bibitem[\protect\citename{Peebles} 1969] {p69}
Peebles P.J.E., 1969, ApJ, 155, 393

\bibitem [\protect\citename{Peebles }1980], 
Peebles P.J.E., 1980, The Large--Scale Structure
of the Universe, Princeton University Press, Princeton

\bibitem [\protect\citename{Quinn \& Binney }1992]{qb92}
Quinn T. \& Binney J., 1992, MNRAS, 255, 729

\bibitem [\protect\citename{Ryden}1988]{ry88}
Ryden B.S., 1988, ApJ, 329, 589

\bibitem[\protect\citename{Shandarin} 1980] {s80}
Shandarin S., 1980, Astrofizika, 16, 769 (1981, Astrophysics, 16, 439)

\bibitem[\protect\citename{Shandarin \& Zel'dovich} 1989] {sz89}
Shandarin S.F. \& Zel'dovich Ya.B. 1989, Rev. Mod. Phys., 61, 185

\bibitem [\protect\citename{van de Weygaert \& Bertschinger }1995]{vdwb95}
van de Weygaert R. \& Bertschinger E., 1996, MNRAS, to be published

\bibitem[\protect\citename{White} 1984] {w84}
White S.D.M., 1984, ApJ, 286, 38

\bibitem[\protect\citename{Zel'dovich} 1970a] {z70a}
Zel'dovich Ya.B., 1970a, A\&A, 5, 84

\bibitem[\protect\citename{Zel'dovich} 1970b] {z70b}
Zel'dovich Ya.B., 1970b, Astrophysics, 6, 164

\end{thebibliography}
\end{document}